\documentclass[11pt]{article}
\pdfoutput=1

\usepackage{jheppub}
\usepackage[latin9,utf8]{inputenc}
\usepackage{latexsym}
\usepackage{revsymb}
\usepackage{amsmath}
\usepackage{amssymb}
\usepackage{graphicx}
\usepackage{epsfig}  
\usepackage{epsf}    
\usepackage{cancel}

\usepackage{subfigure}
\usepackage{amscd}
\usepackage{appendix}
\usepackage{verbatim}
\usepackage{epstopdf}
\usepackage{lineno}

\usepackage{color}

\newcommand{\vev}[1]{\langle #1 \rangle}

\newcommand{\LCP}{\Lambda_{\cancel{\mathcal{CP}}}}

\preprint{IPMU19-0175, OCHA-PP-359}

\title{Higgs Inflation, Vacuum Stability, and Leptogenesis}

\author[1]{Neil D. Barrie,}
\author[2,3]{Akio Sugamoto,}
\author[4]{Tatsu Takeuchi,}
\author[5,6]{and Kimiko Yamashita}

\affiliation[1]{Kavli IPMU (WPI), UTIAS, University of Tokyo, Kashiwa, Chiba 277-8583, Japan}
\affiliation[2]{Department of Physics, Graduate School of Humanities and Sciences, Ochanomizu University, 2-1-1 Otsuka, Bunkyo-ku, Tokyo 112-8610, Japan}
\affiliation[3]{Tokyo Bunkyo SC, Open Universtiy of Japan, Tokyo 112-0012, Japan}
\affiliation[4]{Center for Neutrino Physics, Department of Physics, Virginia Tech, Blacksburg VA 24061, USA}
\affiliation[5]{Institute of High Energy Physics, Chinese Academy of Sciences, Beijing 100049, China}
\affiliation[6]{Department of Physics, National Tsing Hua University, Hsinchu, Taiwan 300}

\emailAdd{neil.barrie@ipmu.jp}
\emailAdd{sugamoto.akio@ocha.ac.jp}
\emailAdd{takeuchi@vt.edu}
\emailAdd{kimiko@ihep.ac.cn}

\abstract{
We consider the introduction of a complex scalar field carrying a global lepton number charge to the Standard Model and the Higgs inflation framework. 
The conditions are investigated under which this model can simultaneously ensure Higgs vacuum stability up to the Planck scale, successful inflation, non-thermal Leptogenesis via the pendulum mechanism, and light neutrino masses. 
These can be simultaneously achieved when the scalar lepton is minimally coupled to gravity, 
that is, when standard Higgs inflation and reheating proceed without the interference of the additional scalar degrees of freedom.
If the scalar lepton also has a non-minimal coupling to gravity, a multi-field inflation scenario is induced, with interesting interplay between the successful inflation constraints and those from vacuum stability and Leptogenesis. The parameter region that can simultaneously achieve the above goals is explored.
}

\arxivnumber{2001.07032}

\begin{document}
\maketitle
\flushbottom

\newpage
\section{Introduction}

In previous publications  we proposed
a model of Leptogenesis in which lepton-number generation 
in the early universe is driven by the oscillation of the inflaton field
during reheating \cite{Takeuchi:2010tm,Bamba:2016vjs,Bamba:2018bwl,Fukugita:1986hr,Linde:2005ht,Dror:2019syi}.
The inflaton $\chi$ is coupled to a scalar-lepton field 
$\phi$ through a $\mathcal{C}$ and $\mathcal{CP}$ violating derivative coupling,
\begin{equation}
\dfrac{g^{\mu\nu}}{\LCP} 
\bigl( \phi^\dagger i\overleftrightarrow{\partial_\mu}\phi \bigr)\partial_\nu\chi\;,
\label{DerivativeCoupling}
\end{equation}
while $\phi$ is subject to a self-coupling potential $V(\phi,\phi^\dagger)$
which contains a lepton number violating term
\cite{Cohen:1987vi} ,
\begin{equation}
V(\phi,\phi^\dagger)
\;=\; -\epsilon_\theta\,\phi^\dagger\phi\,(\phi-\phi^\dagger)^2 + \cdots
\label{Cohen-Kaplan}
\end{equation}
Here, $\epsilon_\theta > 0$ is a dimensionless coupling constant.
The inflaton field $\chi$ oscillates around its potential minimum during reheating, 
transferring its energy to the other degrees of freedom, such that
directed motion in the phase of $\phi$ is generated leading to net lepton number.

The physics involved is analogous to that of a forced pendulum.
The lepton-number violating forces from the potential Eq.~\eqref{Cohen-Kaplan}
provides a restoring force on the phase of $\phi$, pushing it toward one of the periodic potential minima, while the inflaton $\chi$ provides the oscillating external force which violates 
$\mathcal{C}$ and $\mathcal{CP}$.
Energy is transferred from $\chi$ to $\phi$, a clear move away from thermal equilibrium.
When the magnitudes of the two forces and the timing of the `push' from the external force is
just right, the system can enter into a phase-locked state where the `pendulum' rotates
in one direction \cite{Pedersen:1980,D'Humieres:1982}.

In this paper, we investigate whether the Higgs boson $h$
can play the role of the inflaton $\chi$ in our Pendulum Leptogenesis scenario, 
and thereby circumvent the
need to introduce the inflaton as another new scalar field.
We consider a model whose
particle content consists of the Standard Model (SM)
with right-handed neutrinos, and the scalar lepton $\phi$.
This is the minimal particle content in which the Pendulum Leptogenesis
mechanism can be embedded.

The idea of utilizing the SM Higgs boson $h$ as the inflaton was proposed by Bezrukov and Shaposhnikov 
in Ref.~\cite{Bezrukov:2007ep},
and has subsequently been studied by many authors \cite{Bezrukov:2008ut,GarciaBellido:2008ab,Barbon:2009ya,Barvinsky:2009fy,Bezrukov:2009db,Giudice:2010ka,Bezrukov:2010jz,Burgess:2010zq,Lebedev:2011aq,Lee:2018esk,Choi:2019osi}. 
The usual quartic Higgs potential,
\begin{equation}
-\mu_h^2(\Phi^\dagger\Phi) + \lambda_h(\Phi^\dagger\Phi)^2
\quad
\xrightarrow{\;\;\;\Phi=\left(0,h/\sqrt{2}\right)^\mathrm{T}\;\;\;}
\quad
-\dfrac{1}{2}\mu_h^2 h^2 + \dfrac{1}{4}\lambda_h h^4  ~,
\end{equation}
does not have the flatness that is required at large field values
 to realize the slow rolling of the inflaton during inflation \cite{Linde:2007fr}.
This problem is remedied by the introduction of a non-minimal coupling of the
Higgs doublet $\Phi$ to gravity of the form \cite{Bezrukov:2007ep}\footnote{%
For the running of the non-minimal coupling parameter 
from quantum corrections in curved spacetime and its application to inflation,
see, e.g. Refs.~\cite{Muta:1991mw, Buchbinder:1992rb, Mukaigawa:1997nh}.
}
\begin{equation}
\xi\Phi^\dagger \Phi R 
\quad 
\xrightarrow{\;\;\;\Phi=\left(0,h/\sqrt{2}\right)^\mathrm{T}\;\;\;}
\quad
\dfrac{\xi}{2}h^2 R ~,
\end{equation}
where $R$ is the scalar curvature, with $\xi\gg 1$.
This interaction has the effect of flattening the effective inflaton potential at $h/M_p\gg 1/\sqrt{\xi}$.
However, for the Higgs inflation mechanism to work, the Higgs quartic coupling $\lambda_h$,
to which the effective inflaton potential is proportional to, must stay positive up to the Planck scale $M_p$.

Current experimental values of the Higgs and top masses, $m_h$ and $m_t$, suggest that this is not 
the case when considering the SM particle content alone.
Though subject to experimental uncertainties in the top mass $m_t$, the current central SM value is $\lambda_h^{\mathrm{SM}}(M_p) \approx -0.01$, 
which implies that the Higgs vacuum is only metastable within the SM \cite{EliasMiro:2011aa,Degrassi:2012ry,Lebedev:2012sy,Salvio:2013rja,Branchina:2014usa,Bezrukov:2014ipa}.
Indeed, before the discovery of the Higgs, predictions were made on the lower bound of its mass from
the requirement of vacuum stability, which turned out to be too high \cite{Bezrukov:2008ej,Bezrukov:2009db,Allison:2013uaa}.
New particle contributions to the renormalization group equation (RGE) of $\lambda_h$ are necessary to render
$\lambda_h(M_p)$ positive.
This requires new physics to couple to the Higgs, 
with a possible solution suggested by the Higgs portal models\footnote{The Higgs boson is the ``portal'' to new physics, e.g. the dark sector.} \cite{Patt:2006fw}.
For instance, Refs.~\cite{Lebedev:2011aq,Lebedev:2012zw,Ema:2017ckf} study a model in which the Higgs is coupled to a new SM singlet scalar which has the desired effect. The scalar lepton in our model could also function in this capacity.

Thus, our model can lead to successful Higgs inflation, Leptogenesis during reheating, and stability of the Higgs vacuum.\footnote{%
Possible connections between Leptogenesis and Higgs vacuum stability have been considered, e.g., in Refs.~\cite{Salvio:2015cja, Ipek:2018sai,Croon:2019dfw}.
}
The question is what parameter range can  all three  be accomplished simultaneously.
This paper is structured as follows: 
In Section~\ref{Model} the framework of the model is introduced with descriptions of the scalar sector of our model. 
Section~\ref{Dym_mod} provides a summary of the standard Higgs Inflation scenario and the dynamics of the Pendulum Leptogenesis mechanism. 
In Section~\ref{Stability}, we analyze the conditions for vacuum stability in our model with particular reference to the utilization of the scalar threshold effect. 
In Section~\ref{Revisit}, the inflationary scenario is explored in the case of a non-minimal coupling of the scalar lepton to gravity, and consistency with the vacuum stability constraints is confirmed. The dynamics of the reheating epoch is also discussed and the requirements for successful Pendulum Leptogenesis considered. 
Finally, in Section~\ref{Conclusion} we conclude with a discussion of the results and implications.

\section{Description of the Model}
\label{Model}

\subsection{Particle Content}

In addition to the SM plus the three generations of right-handed neutrinos, we introduce
a complex scalar field $\phi$ which carries lepton number 2.
We identify lepton number as the charge under the global $U(1)$ transformation,
\begin{equation}
L\;\to\;e^{i\alpha}L\;,\quad
\ell_R\;\to\;e^{i\alpha}\ell_R\;,\quad
\nu_R\;\to\;e^{i\alpha}\nu_R\;,\quad
\phi\;\to\;e^{2i\alpha}\phi\;,
\end{equation}
where $L=(\nu_L,\ell_L)^\mathrm{T}$ is the left-handed lepton doublet.
The lepton-number current carried by the scalar-lepton $\phi$ is,
\begin{equation}
j_L^\mu \;=\; 2
\left(\phi^\dagger i\overleftrightarrow{\partial^\mu}\phi\right)
\;.
\end{equation}

\subsection{Interactions among Particles -- Dimension 4 Operators}

The SM particles interact among themselves in the usual way.
We assume that $\phi$ and the right-handed neutrinos $\nu_R$ couple to the SM fields in a lepton number preserving form:
\begin{equation}
\mathcal{L}_{\phi,\nu_R} \;=\; 
\left(\,
g\,\phi^* \overline{\nu_R^c}\nu_R^{\phantom{\dagger}} 
\,+\, y\,\overline{L}\,\widetilde{\Phi} \nu_R^{\phantom{\dagger}} 
\,\right)
\,+\, \mathrm{h.c.}
\label{nu_int}
\end{equation}
where $\Phi$ is the Higgs doublet, and $\widetilde{\Phi}=i\sigma_2\Phi^*$.
The $\phi^* \overline{\nu_R^c}\nu_R^{\phantom{\dagger}}$ interaction will allow
the scalar lepton $\phi$ to decay into $\nu_R\nu_R$ pairs, 
converting the lepton number carried by the scalar $\phi$ into fermionic lepton number.
Once $\phi$ and $\Phi$ develop 
vacuum expectation values, the interactions will generate the right-handed Majorana and Dirac masses for the neutrinos.
Upon diagonalization of the resulting mass matrix, we obtain 
light Majorana masses for the active neutrinos via the seesaw mechanism \cite{Yanagida:1979as,Ramond:1979py}.

The scalar potential involving both $\phi$ and $\Phi$ is taken to be,
\begin{eqnarray}
V(\Phi,\Phi^\dagger,\phi,\phi^\dagger) 
& = & 
\lambda_h\biggl(\dfrac{v_h^2}{2}-\Phi^\dagger\Phi\biggr)^2
+ \lambda_\phi\biggl(\dfrac{v_\varphi^2}{2}-\phi^\dagger\phi\biggr)^2
\cr
& & 
+ \kappa\,\biggl(\dfrac{v_h^2}{2}-\Phi^\dagger\Phi\biggr)\biggl(\dfrac{v_\varphi^2}{2}-\phi^\dagger\phi\biggr)
- \epsilon_{\theta}\phi^\dagger\phi\left(\phi-\phi^\dagger\right)^2
\;.
\end{eqnarray}
Note that the $\epsilon_\theta$ term violates lepton number but preserves $\mathcal{C}$ and $\mathcal{CP}$.
The $\kappa$ term couples the Higgs doublet $\Phi$ to the scalar-lepton $\phi$,
which is a SM singlet, and is a type of \textit{Higgs portal} interaction ~\cite{Lebedev:2011aq,Lebedev:2012zw}.
We consider the unitary gauge $\Phi = \left(0,h/\sqrt{2}\right)^\mathrm{T}$, and rewrite $\phi$ as,
\begin{equation}
\phi \;=\; \dfrac{1}{\sqrt{2}}\,\varphi\,e^{i\theta}\;,
\end{equation}
where $\varphi$ and $\theta$ are real fields that transform as,
\begin{equation}
\varphi \;\to\; \varphi\;,\qquad
\theta \;\to\; \theta + 2\alpha\;,
\end{equation}
under the lepton-number $U(1)$ transformation $\phi\to e^{2i\alpha}\phi$.
The scalar potential becomes,
\begin{eqnarray}
V(\Phi,\Phi^\dagger,\phi,\phi^\dagger) \quad\to\quad
V(h,\varphi,\theta) 
& = & \dfrac{\lambda_h}{4}\left(v_h^2-h^2\right)^2 
+ \dfrac{\lambda_\phi}{4}\left(v_\varphi^2-\varphi^2\right)^2 \cr
& & 
+ \dfrac{\kappa}{4}\left(v_h^2-h^2\right)\left(v_\varphi^2-\varphi^2\right)
+ \epsilon_\theta\,\varphi^4\sin^2\theta\;.
\label{scalar-potential-V}
\end{eqnarray}
The loss of translational invariance in $\theta$, due to the $\epsilon_\theta$ term, breaks
lepton-number conservation.
The potential minima are at,
\begin{equation}
\vev{h^2}\;=\;v_h^2\;,\qquad 
\vev{\varphi^2}\;=\;v_\varphi^2\;,\qquad 
\vev{\theta}\;=\;n\pi \;,\quad (n\in\mathbb{Z})
\;,
\label{VEVs}
\end{equation}
provided that,
\begin{equation}
\lambda_h\,>\,0\;,\quad
\lambda_\phi \,>\,0\;,\quad
\epsilon_\theta \,>\,0\;,\quad
4\lambda_h\lambda_\phi \,>\, \kappa^2\;.
\label{Vstability}
\end{equation}
The last condition on $\kappa$ will prove important later.
The masses of the oscillations around these minima are,
\begin{equation}
m_h^2 \;=\; 2\lambda_h v_h^2\;,\qquad
m_\varphi^2 \;=\; 2\lambda_\phi v_\varphi^2\;,\qquad
m_\theta^2 \;=\; 2\epsilon_\theta v_\varphi^2\;.
\label{ScalarMasses}
\end{equation}
The lepton-number current carried by $\phi$ in the $\varphi$ and $\theta$ variables is,
\begin{equation}
j_L^\mu \;=\; -2\,\varphi^2
\partial^\mu\theta
\;,
\end{equation}
and the lepton-number density is $j_L^0 \;=\; -2\varphi^2\dot{\theta}$.
Thus, to generate lepton number from the dynamics of $\phi$, we need both $\varphi\neq 0$ and $\dot{\theta}\neq 0$~.

\subsection{Non-Minimal Interactions with Gravity}

Coupling the particle content to gravity, we obtain the action
\begin{eqnarray}
\lefteqn{S_J 
\;=\; \int d^4x\sqrt{-g}
\Biggl[\,-\dfrac{M_p^2}{2}R -\xi\Phi^\dagger\Phi R -\zeta \phi^\dagger\phi\,R
+ g^{\mu\nu}\,\partial_\mu\Phi^\dagger\,\partial_\nu\Phi
+ g^{\mu\nu}\,\partial_\mu\phi^\dagger\,\partial_\nu\phi
}\cr
& & \qquad\qquad
-V(\Phi,\Phi^\dagger,\phi,\phi^\dagger) 
+ \mathcal{L}_{\phi,\nu_R} + \cdots
\Biggr] + \cdots
\cr
& & \vphantom{|}\cr
& \xrightarrow{\mbox{$U$ gauge}} & \int d^4x\sqrt{-g}
\Biggl[\,-\dfrac{M_p^2}{2}\left(1 + \dfrac{\xi h^2}{M_p^2} + \dfrac{\zeta \varphi^2}{M_p^2}\right)R 
+ \dfrac{1}{2} g^{\mu\nu}\,\partial_\mu h\,\partial_\nu h
+ \dfrac{1}{2} g^{\mu\nu}\,\partial_\mu\varphi\,\partial_\nu\varphi
\cr
& & \qquad\qquad\qquad
+ \dfrac{\varphi^2}{2}\,g^{\mu\nu}\,\partial_\mu\theta\,\partial_\nu\theta
-V(h,\varphi,\theta) 
+ \mathcal{L}_{\phi,\nu_R} + \cdots
\Biggr] + \cdots
\;,
\end{eqnarray}
where $M_p = 1/\sqrt{8\pi G} = 2.4\times 10^{18}\,\mathrm{GeV}$
is the reduced Planck mass,
and the ellipses represent contributions of SM particles and interactions that are not shown explicitly,
and surface terms \cite{York:1972sj,Gibbons:1976ue}.
Note that for scalars, the covariant derivative $\nabla_\mu$ and the partial derivative $\partial_\mu$
coincide.
In addition to the non-minimal coupling of the Higgs doublet $\Phi$ to gravity, $\xi\Phi^\dagger\Phi R$, we also include a non-minimal coupling of the scalar-lepton $\phi$ to gravity, $\zeta\phi^\dagger\phi R$. 
We first consider the case $\zeta=0$, and then generalize to the case $\zeta\neq 0$.

\subsection{$\mathcal{C}$ and $\mathcal{CP}$ violation -- Dimension 6 Operator}

We introduce a dimension-6 operator of the form
\begin{equation} 
\mathcal{L}_{\cancel{\mathcal{CP}}}(\Phi,\Phi^\dagger,\phi,\phi^\dagger)
\;=\;
\dfrac{2g^{\mu\nu}}{\Lambda^2}\,
(\phi^\dagger i\overleftrightarrow{\partial_\mu}\phi)\,\partial_\nu(\Phi^\dagger \Phi)
\;,
\label{Dim6op}
\end{equation}
which changes sign under $\mathcal{C}$ and $\mathcal{CP}$ transformations.\footnote{%
A possible method of inducing this interaction is including the following term in the
Jordan frame action:
\[
S_{\cancel{\mathcal{CP}}} \;=\; 
\int d^4x\sqrt{-g}\;
\left[
g^{\mu\nu}
\nabla_\mu (\phi^\dagger\, i\!\overleftrightarrow{\nabla}\!\!{}_\nu\,\phi)
\right]
\;.
\]
This is an integral of the divergence of a non-conserved current, and as such will not vanish.
In the Einstein frame, this becomes
\[
S_{\cancel{\mathcal{CP}}} \;=\; 
\int d^4x\sqrt{-\tilde{g}}\;
\left[
\dfrac{\tilde{g}^{\mu\nu}}{\Omega^2}
\nabla_\mu (\phi^\dagger\, i\!\overleftrightarrow{\nabla}\!\!{}_\nu\,\phi)
\right]
\;,
\]
and expanding $\Omega^2$ in powers of $\xi h^2/M_p^2$ will provide the required interaction after partial integration.
}
This interaction leads to a derivative coupling between $\phi$ and the effective inflaton
$\chi\sim h^2$ during reheating:
\begin{equation}
\mathcal{L}_{\cancel{\mathcal{CP}}}(\Phi,\Phi^\dagger,\phi,\phi^\dagger)
\quad\to\quad
\mathcal{L}_{\cancel{\mathcal{CP}}}(h,\varphi,\theta)
\;=\;
-\frac{g^{\mu\nu}}{\Lambda^2}
\,\varphi^2 (\partial_\mu\theta)\left[\partial_\nu(h^2)\right]
\;.
\label{CPterm}
\end{equation}
The strength of this interaction is adjusted by the choice of scale $\Lambda$.

\section{Dynamics of the Model}
\label{Dym_mod}
\subsection{Inflation: $\zeta=0$ Case}
\label{Higgs_inf}

We begin the analysis of our model by first setting $\zeta=0$ for the sake of simplicity,
so that only the dynamics of the Higgs $h$ is involved in inflation.
Consequently, the content of this subsection will be a review of the standard Higgs inflation scenario, 
so readers who are familiar with the topic may skip to Section~\ref{Reheating}.

\begin{figure}[t]
\begin{center}
\subfigure[Higgs-Inflaton relation]{\includegraphics[height=6.5cm]{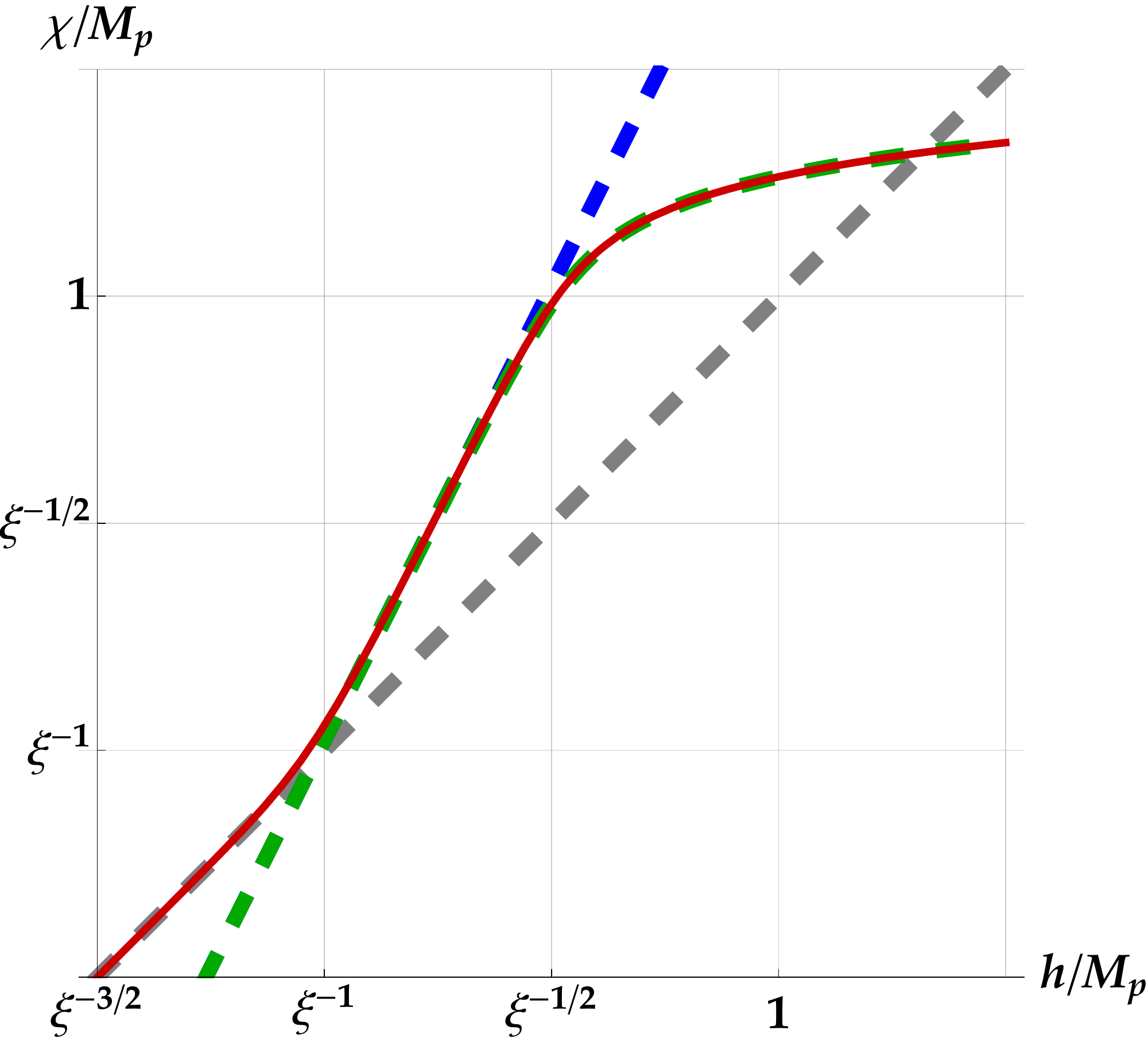}}
\subfigure[Effective Inflaton Potential]{\includegraphics[height=6.5cm]{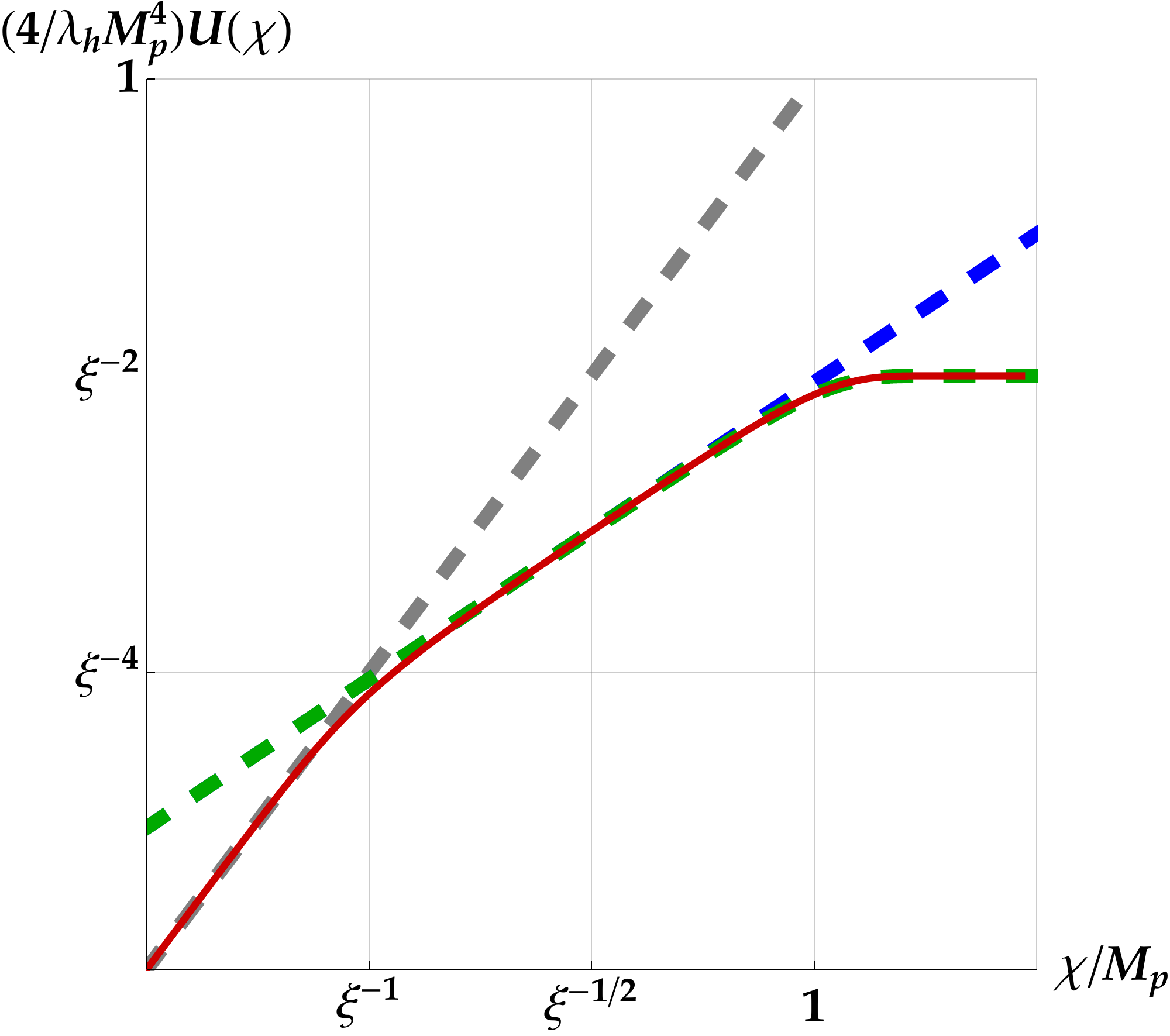}}
\caption{(a) The relation between the effective inflaton field $\chi$ and the SM Higgs field $h$ (red solid line).
In the region $h/M_p\ll\xi^{-1}$, the relation implies $\chi \approx h$ (grey dashed line).
In the region $\xi^{-1}<h/M_p$, the relation is well approximated by
$\chi/M_p \approx \sqrt{3/2}\ln[1+\xi(h/M_p)^2]$ (green dashed line),
which in the subregion $\xi^{-1}<h/M_p<\xi^{-1/2}$ is further approximated by
$\chi/M_p \approx \sqrt{3/2}\,\xi(h/M_p)^2$ (blue dashed line).
(b) The effective potential $U(\chi)=U_0(h)/\Omega^4(h)$ (red solid line), 
where $U_0(h)=\lambda_h h^4/4$ and $\Omega^2(h)=1+\xi h^2/M_p^2$.
In the region $\chi/M_p < \xi^{-1}$ the potential behaves as $U(\chi)\sim \chi^4$ (grey dashed line).
For $\xi^{-1} < \chi/M_p$, the potential matches the behaviour of the Starobinsky potential
(green dashed line), which behaves as $U(\chi)\sim \chi^2$ (blue dashed line) in the range
$\xi^{-1}<\chi/M_p<1$.
See also Figures 1 and 2 of Ref.~\cite{Bezrukov:2008ut}.
}
\label{chi-h}
\end{center}
\end{figure}

The parts of the action relevant for Higgs inflation are \cite{Bezrukov:2007ep},
\begin{eqnarray}
S_J & = & \int d^4x \,\sqrt{-g}\,
\biggl[\, -\frac{M_p^2}{2}\left(1+\frac{\xi h^2}{M_p^2}\right)R+
\frac{1}{2} g^{\mu\nu}\,\partial_\mu h\,\partial_\nu h -U_0(h)
+ \cdots
\,\biggr]
\;,
\label{higgs}
\end{eqnarray}
where it is assumed that $\xi\gg 1$, and,
\begin{equation}
U_0(h) 
\;=\; \dfrac{\lambda_h}{4}\left(v_h^2-h^2\right)^2
\quad\xrightarrow{h^2\gg v_h^2}\quad \dfrac{\lambda_h}{4}h^4\;,
\end{equation}
is the usual Higgs potential.
In order to manifest  
the inflationary dynamics, we perform a conformal transformation 
from the Jordan frame to the Einstein frame \cite{Faulkner:2006ub,Bezrukov:2007ep},
\begin{equation}
g_{\mu\nu} \;\to\; \tilde{g}_{\mu\nu} \,=\, \Omega^2 g_{\mu\nu}\;,\quad
g^{\mu\nu} \;\to\; \tilde{g}^{\mu\nu} \,=\, \dfrac{1}{\Omega^{2}}\,g^{\mu\nu}\;,
\label{ConformalTransform}
\end{equation}
with,
\begin{equation}
\Omega^2 \,=\, \left(1 + \dfrac{\xi h^2}{M_p^2}\right)\;.
\end{equation}
Note that the determinant of the metric $g=\det(g_{\mu\nu})$ scales as,
\begin{equation}
\sqrt{-g} \;=\; \dfrac{\sqrt{-\tilde{g}}}{\Omega^4}\;,
\end{equation}
while the scalar curvature in the Einstein frame $\tilde{R}$ is related to that in the Jordan frame $R$ via
(see, e.g., Appendix D of \cite{Wald:1984rg}, also \cite{Faraoni:1998qx})
\begin{eqnarray}
\tilde{R} & = & 
\dfrac{1}{\Omega^{2}}
\left(R-\dfrac{6 g^{\mu\nu}\nabla_\mu\nabla_\nu\Omega}{\Omega}\right)
\cr
& \downarrow & \cr
\dfrac{R}{\Omega^2}
& = & 
\tilde{R}
- \tilde{g}^{\mu\nu}\dfrac{3(\partial_\mu\Omega^2)(\partial_\nu\Omega^2)}{2\Omega^4}
+ \mbox{(total derivative)}~,
\end{eqnarray}
and the action is rendered into the form,
\begin{eqnarray}
S_E  & = & \int d^4x \,\sqrt{-\tilde{g}}\,
\Biggl[\,-\frac{M_p^2}{2}\,\tilde{R}
\,+\,\frac{\tilde{g}^{\mu\nu}}{2}
\left\{
M_p^2\dfrac{3(\partial_\mu\Omega^2)(\partial_\nu\Omega^2)}{2\Omega^4} 
+\dfrac{1}{\Omega^2}\,\partial_\mu h\,\partial_\nu h 
\right\}
\,-\,\frac{1}{\Omega^4}\,U_0(h)
\,+\,\cdots
\,\Biggr]
\;,\cr
& &
\label{higgs_E}
\end{eqnarray}
which decouples $\tilde{R}$ from $h$.
The Einstein frame inflaton field $\chi$ is defined via,
\begin{equation}
M_p^2\dfrac{3(\partial_\mu\Omega^2)(\partial_\nu\Omega^2)}{2\Omega^4} 
+\dfrac{1}{\Omega^2}\,\partial_\mu h\,\partial_\nu h 
\;=\; \dfrac{(6\xi^2 h^2/M_p^2)+\Omega^2}{\Omega^4}\,\partial_\mu h\,\partial_\nu h
\;\equiv\; \partial_\mu\chi\,\partial_\nu\chi
\;,
\label{chi-def}
\end{equation}
that is, $\chi$ is obtained from $h$ via the integration of,
\begin{equation}
\frac{d\chi}{dh}
\;=\; \dfrac{\sqrt{(6\xi^2 h^2/M_p^2)+\Omega^2}}{\Omega^2}
\;.
\end{equation}
%
%
%
Note that for very large field values, $h \gg M_p/\sqrt{\xi}$, we have $\Omega^2 = 1 + (\xi h^2/M_p^2) \gg 1$, and consequently,
\begin{equation}
\partial_\mu\chi\,\partial_\nu\chi \;\approx\; M_p^2\dfrac{3(\partial_\mu\Omega^2)(\partial_\nu\Omega^2)}{2\Omega^4} 
\qquad\to\qquad
\chi\;\approx\; \sqrt{\dfrac{3}{2}}M_p\ln\Omega^2\;.
\label{chi-inflation}
\end{equation}
In this regime, all kinetic terms not explicitly shown in Eq.~\eqref{higgs_E} are suppressed by $1/\Omega^2$.
For small field values, $h \ll M_p/\xi$, we have $(\xi^2 h^2/M_p^2) \ll \Omega^2 \approx 1$, and consequently,
\begin{equation}
\partial_\mu\chi\,\partial_\nu\chi \;\approx\; \partial_\mu h\,\partial_\nu h
\qquad\to\qquad
\chi\;\approx\; h\;.
\label{chi-after-reheating}
\end{equation}
In the intermediate range, $M_p/\xi \ll h \ll M_p/\sqrt{\xi}$, we can expand,
\begin{equation}
\chi\;\approx\; \sqrt{\dfrac{3}{2}}M_p\ln\Omega^2
\;=\; \sqrt{\dfrac{3}{2}}M_p\ln\left(1+\dfrac{\xi h^2}{M_p^2}\right)
\;\approx\; \sqrt{\dfrac{3}{2}}\dfrac{\xi h^2}{M_p} ~.
\label{chi-reheating}
\end{equation}
Therefore, the relation between $h$ and $\chi$ has three key regimes which correspond to different cosmological epochs in the inflationary scenario, namely:
\begin{equation}
\dfrac{\chi}{M_p}
\approx
\left\{
\begin{array}{lll}
\dfrac{h}{M_p}
& \mbox{for $\dfrac{h}{M_p} \ll\dfrac{1}{\xi}$} 
& \mbox{(after reheating)}
\\
\sqrt{\dfrac{3}{2}}\,\xi\left(\dfrac{h}{M_p}\right)^2 
& \mbox{for $\dfrac{1}{\xi}\ll \dfrac{h}{M_p} \ll \dfrac{1}{\sqrt{\xi}}$\quad} 
& \mbox{(reheating)}
\\
\sqrt{\dfrac{3}{2}} \ln\Omega^2 =
\sqrt{\dfrac{3}{2}} \ln\left[1+\xi\left(\dfrac{h}{M_p}\right)^2\right] \quad
& \mbox{for $\dfrac{1}{\sqrt{\xi}}\ll \dfrac{h}{M_p}$} 
& \mbox{(inflation)}
\end{array}
\right.
\label{chi-h-relation}
\end{equation}
where the duration of the intermediate reheating period 
is dictated by the size of the non-minimal coupling $\xi$.
See Figure~\ref{chi-h}(a).
Note that
$h/M_p\sim 1/\xi$ corresponds to $\chi/M_p\sim 1/\xi$, whereas 
$h/M_p\sim 1/\sqrt{\xi}$ corresponds to $\chi/M_p\sim 1$.
Replacing $h$ with $\chi$, 
the Einstein frame action is,
\begin{equation}
S_E \;=\; \int d^4x \,\sqrt{-\tilde{g}}\,
\left[\, -\frac{M_p^2}{2}\tilde{R}+
\frac{1}{2}\tilde{g}^{\mu\nu} \partial_\mu \chi \,\partial_\nu \chi - U(\chi)
+ \cdots
\,\right]
\;,
\label{higgs_E_chi}
\end{equation}
where $U(\chi) = U_0(h)/\Omega^4$.
The factor of $\Omega^4$ in the denominator leads to the flattening out of $U_0(h)/\Omega^4$ for large field values:
\begin{equation}
\dfrac{1}{\Omega^4}U_0(h) 
\;\xrightarrow{h\gg v_h}\; \dfrac{\lambda_h}{4}\dfrac{h^4}{\bigl[1+(\xi h^2/M_p^2)\bigr]^2}
\;=\; \frac{\lambda_h M_p^4}{4\,\xi^2}\left(1-\dfrac{1}{\Omega^2}\right)^2
\;\xrightarrow{h\gg M_p/\sqrt{\xi}}\; 
\frac{\lambda_h M_p^4}{4\,\xi^2}
\;.
\end{equation}
In terms of $\chi$, 
the effective potential $U(\chi)$ takes on the forms,
\begin{equation}
U(\chi) \approx
\left\{
\begin{array}{lll}
\dfrac{1}{4}\lambda_h\chi^4 
& \mbox{for $\dfrac{\chi}{M_p}\ll\dfrac{1}{\xi}$}
& \mbox{(after reheating)}
\\
\dfrac{1}{2} \mu_{\inf}^2 \chi^2 
& \mbox{for $\dfrac{1}{\xi} \ll \dfrac{\chi}{M_p} \ll 1$\quad}
& \mbox{(reheating)}
\\
\dfrac{3}{4}\mu_{\inf}^2 M_p^2
\left[1-e^{-\sqrt{\frac{2}{3}}(\chi/M_p)}\right]^2 \quad
& \mbox{for $1\ll \dfrac{\chi}{M_p}$}
& \mbox{(inflation)}
\end{array}
\right.
\label{Uchi}
\end{equation}
where 
\begin{equation}
\mu_{\inf}^2 \;=\; \dfrac{\lambda_h M_p^2}{3\,\xi^2}\;.
\label{mu_Higgs_inf}
\end{equation}
See Figure~\ref{chi-h}(b).
Note that in the inflationary regime, the potential is analogous to the Starobinsky $R^2$ inflation scenario \cite{Starobinsky:1980te,Whitt:1984pd,Jakubiec:1988ef,Maeda:1988ab,Barrow:1988xh,Faulkner:2006ub,Bezrukov:2011gp}.

\bigskip
\subsection{Characteristics of the Reheating Epoch}
\label{Reheating}

The Starobinsky model is known to provide a good fit to the Planck data \cite{Akrami:2018odb}.
The value of the mass scale $\mu_{\inf}$ in the Starobinsky potential is constrained by the scalar power spectrum amplitude $A_s$ of the Cosmic Microwave Background (CMB) to be \cite{Faulkner:2006ub,Aldabergenov:2018qhs}:
\begin{equation}
\mu_{\inf} \;\approx\; 3\times 10^{13}\,\textrm{GeV}\;.
\label{mu_from_CMB}
\end{equation}
For our inflation model, in which we have Eq.~\eqref{mu_Higgs_inf}, this requires,
\begin{equation}
\frac{\lambda_h}{\xi^2} 
\;=\; \dfrac{3\mu_{\inf}^2}{M_p^2} \;\approx\; 5\times 10^{-10}\;.
\label{lambda_over_xi2}
\end{equation}
If we use the tree level-value of $\lambda_h$, 
\begin{equation}
\lambda_h \;=\; \dfrac{m_h^2}{2v_h^2} \;=\; \dfrac{G_F m_h^2}{\sqrt{2}} \;\approx\; 0.13\;,
\label{TreeLambdah}
\end{equation}
we find,
\begin{equation}
\xi \;\approx\; 2\times 10^4\;.
\end{equation}
This large value of $\xi$ is favourable for the Leptogenesis component of our model, since
it provides a long reheating period in which the Pendulum Leptogenesis mechanism can take place.
Indeed, during the reheating period  the inflaton $\chi$ evolves from $M_p$ to
$M_p/\xi$, and the Hubble rate $H=\dot{a}/a$ evolves in the range,
\begin{eqnarray}
H 
\,=\, \sqrt{\dfrac{\rho}{3M_p^2}} 
\,\approx\, \sqrt{\dfrac{\mu_{\inf}^2\chi^2}{6M_p^2}}
\;\;
& = & \;\;\dfrac{\mu_{\inf}}{\sqrt{6}} \;\;\to\;\; \dfrac{\mu_{\inf}}{\sqrt{6}\xi}\cr
& \simeq & \;\;10^{13}\,\mathrm{GeV} \;\;\to\;\; 6\times 10^{8}\,\mathrm{GeV}\;.\vphantom{\Bigg|}
\label{ReheatingHubbleRange}
\end{eqnarray}
The Hubble rate at the end of the reheating period, $H_\mathrm{reh}\simeq 6\times 10^{8}\,\mathrm{GeV}$, gives the reheating temperature:
\begin{eqnarray}
3M_p^2 H_\mathrm{reh}^2 
& = & \rho_{\mathrm{rad,reh}} 
\,=\, g_{*}\dfrac{\pi^2}{30}T_\mathrm{reh}^4 
\cr
& \downarrow & \cr
T_\mathrm{reh}
& = &
\left(\dfrac{15}{g_{*}}\right)^{1/4}\left(\dfrac{M_p\mu_{\inf}}{\pi\xi}\right)^{1/2}
\,\simeq\,\dfrac{(3\times 10^{15}\,\mathrm{GeV})}{\sqrt{\xi}}
\quad\xrightarrow{\xi=2\times 10^4}\quad 2\times 10^{13}\,\mathrm{GeV}\;,
\cr
& & 
\label{ReheatingTemp}
\end{eqnarray}
where $g_{*}=106.75$ is the effective number of massless degrees of freedom \cite{Kolb:1990vq,Husdal:2016haj}.
As will be shown in the next section,
lepton asymmetry generation via the Pendulum mechanism can occur at approximate Hubble rates of,
\begin{equation}
H_d 
\;\simeq\; (3n\times 10^{-6})\LCP
\;,
\end{equation}
%
where $n\in\mathbb{N}$ and
$\LCP$ is the $\mathcal{C}$ and $\mathcal{CP}$ breaking scale, cf. Eq.~\eqref{DerivativeCoupling},
when considering an inflaton $\chi$ oscillating in an 
$\frac{1}{2}\mu_{\inf}^2 \chi^2$ potential during reheating. 
If $\LCP = \Lambda_{\mathrm{GUT}} = 10^{16}\,\mathrm{GeV}$, we have,
\begin{equation}
H_d \;\simeq\; 3n\times 10^{10}\,\mathrm{GeV}.
\end{equation}
Comparison with Eq.~\eqref{ReheatingHubbleRange} indicates that there are multiple opportunities
during reheating for Pendulum Leptogenesis to occur.
Thus, the characteristics of the reheating epoch in the Higgs inflation framework 
matches the conditions necessary to host Pendulum Leptogenesis dynamics.

\subsection{Pendulum Leptogenesis}
\label{Pend}

We continue our analysis in the Einstein frame.
In contrast to the inflationary epoch when $\Omega^2\gg 1$, 
during the reheating epoch we have $\Omega^2 \gtrapprox 1$,
and the dynamics of all the non-inflaton fields become important as energy is transferred from the
inflaton to those fields.

\subsubsection*{Derivative Coupling Between the Inflaton and Scalar-Lepton}

Recall that we introduced the $\mathcal{C}$ and $\mathcal{CP}$ violating dimension-6 operator
in Eq.~\eqref{CPterm}.
During the reheating epoch we have $\xi (h/M_p)^2 \sim \sqrt{2/3}(\chi/M_p)$, 
which induces the derivative coupling between the lepton-number current and the inflaton $\chi$:
\begin{equation}
\mathcal{L}_{\cancel{\mathcal{CP}}}
\;=\;
-\frac{\tilde{g}^{\mu\nu}}{\Lambda^2}
\,\varphi^2 (\partial_\mu\theta)\left[\partial_\nu(h^2)\right]
\;\approx\; -\sqrt{\dfrac{2}{3}}\,\dfrac{M_p}{\xi\Lambda^2}
\,\tilde{g}^{\mu\nu}\varphi^2(\partial_\mu\theta)(\partial_\nu\chi)
\;=\; -\dfrac{\tilde{g}^{\mu\nu}}{\LCP}
\,\varphi^2(\partial_\mu\theta)(\partial_\nu\chi)
\;,
\label{CP_term}
\end{equation}
where we have set,
\begin{equation}
\LCP
\;\equiv\; \sqrt{\dfrac{3}{2}}\dfrac{\xi\Lambda^2}{M_p}
\;,
\end{equation}
cf. Eq.~\eqref{DerivativeCoupling}.

\subsubsection*{Further Simplifications and the Sakharov Conditions}

Including the above term, the relevant terms in the action during reheating are,
\begin{eqnarray}
S 
& = & \int d^4x \,\sqrt{-\tilde{g}}\,
\Biggl[\,
\frac{1}{2}\,\tilde{g}^{\mu\nu}\,\partial_\mu\chi\,\partial_\nu\chi
+ \frac{1}{2}\,\tilde{g}^{\mu\nu}\,\partial_\mu\varphi\,\partial_\nu\varphi
+ \dfrac{\varphi^2}{2}\tilde{g}^{\mu\nu}\,\partial_{\mu}\theta\partial_{\nu}\theta
\cr
& & \qquad\qquad\qquad 
-\dfrac{1}{2}\mu_{\inf}^2\chi^2 
-\dfrac{\lambda_\phi}{4}(v_\varphi^2-\varphi^2)^2
-\epsilon_\theta\varphi^4\sin^2\theta 
-\dfrac{\tilde{g}^{\mu\nu}}{\LCP}
\,\varphi^2(\partial_\mu\theta)(\partial_\nu\chi)
\Biggr]\;,
\cr
& &
\label{Ch3eq:actiona}
\end{eqnarray}
where for the moment we have dropped the $h^2\varphi^2$ portal coupling $\kappa$ in $V(h,\varphi,\theta)$.
The $\mathcal{CP}$ violation scale $\LCP$ will be determined later.

We further assume for simplicity that $m_\varphi = \sqrt{2\lambda_\phi}v_\varphi \gg \mu_{\inf}$, so that $\varphi\approx v_\varphi$
during reheating and that its dynamics need not be considered.
(This assumption will be relaxed later.)
The action becomes
\begin{eqnarray}
S 
& = & \int d^4x \,\sqrt{-\tilde{g}}\,
\Biggl[\,
\frac{1}{2}\,\tilde{g}^{\mu\nu}\,\partial_\mu\chi\,\partial_\nu\chi
+ \dfrac{v_\varphi^2}{2}\tilde{g}^{\mu\nu}\,\partial_{\mu}\theta\partial_{\nu}\theta
\cr
& & \qquad\qquad\qquad 
-\dfrac{1}{2}\mu_{\inf}^2\,\chi^2 
-\epsilon_\theta v_\varphi^4\sin^2\theta 
-\dfrac{\tilde{g}^{\mu\nu}}{\LCP}
\,v_\varphi^2(\partial_\mu\theta)(\partial_\nu\chi)
+ \cdots
\Biggr]\;.
\label{Ch3eq:actionb}
\end{eqnarray}
We take $\tilde{g}_{\mu\nu}$ to be the flat Friedmann-Robertson-Walker metric with scale factor $a(t)$.
Given this isotropic and homogeneous background, we extend this assumption to the properties of the scalar lepton and inflaton
for which spatial variation will be ignored in our analysis. 
Therefore, in this parametrization the action takes the form,
\begin{equation}
S \;=\; 
\int d^4x  \; a(t)^3
\left[\,
\frac{1}{2}\,\dot{\chi}^2
\,-\, \frac{1}{2}\,\mu_{\inf}^2\,\chi^2 
\,+\, \frac{v_\varphi^2}{2}\,\dot{\theta}^2
\,-\, \epsilon_{\theta}\,v_\varphi^4 \sin^2\theta
\,-\, \frac{v_\varphi^2}{\LCP}\,\dot{\theta}\,\dot{\chi} 
\,\right]\;.
\label{Ch3eq:action2a2}
\end{equation}
This action compactly showcases the main ingredients of our model. 
Note how the Sakharov conditions \cite{Sakharov:1967dj} are satisfied: 
Firstly, $L$ violation is achieved by the potential $\epsilon_{\theta}\varphi^4 \sin^2\theta$,
which breaks the translational invariance in $\theta$.
Secondly, the derivative coupling between $ \theta $ and $ \chi $ provides $\mathcal{C}$ and $\mathcal{CP}$ violation. Lastly, the required push out-of-thermal-equilibrium is provided by the reheating epoch, induced by the coherent oscillation of the inflaton field $\chi$. 
The generated scalar-lepton-number asymmetry will be converted to  fermions via the decay of the scalar lepton,
and  later redistributed into a net baryon number by the action of the $B-L$ conserving sphaleron processes \cite{Klinkhamer:1984di,Kuzmin:1985mm,Trodden:1998ym,Sugamoto:1982cn}.

\subsubsection*{Behaviour of the Inflaton}

To determine the conditions under which driven motion can be generated within this framework, we must first specify the inflaton dynamics. We wish for the inflaton's motion to be unaffected by the dynamics of $\theta$. This is to ensure that the properties of the reheating epoch and the coherent oscillation of the inflaton are retained. 

The equations of motion obtained from the action, Eq.~\eqref{Ch3eq:action2a2}, by varying $\chi$ and $\theta$ are,
\begin{eqnarray}
& &
\Bigl(\ddot{\chi}+3H\dot{\chi}\Bigr)
-\dfrac{v_\varphi^2}{\LCP}
 \Bigl(\ddot{\theta}+3H\dot{\theta}\Bigr)
+\mu_{\inf}^2\chi \;=\; 0\;,
\label{EoM-nophi-chi1}
\vphantom{\Bigg|}\\
& & 
\Bigl(\ddot{\theta}+3H\dot{\theta}\Bigr)
-\dfrac{1}{\LCP}
 \Bigl(\ddot{\chi}+3H\dot{\chi}\Bigr)
+\epsilon_\theta v_\varphi^2 \sin 2\theta
\;=\; 0\;,
\label{EoM-nophi-theta1}
\vphantom{\Bigg|}
\end{eqnarray}
which can be rearranged to yield,
\begin{eqnarray}
& & \biggl(1-\dfrac{v_\varphi^2}{\LCP^2}\biggr)
\Bigl(\ddot{\chi}+3H\dot{\chi}\Bigr) 
+\mu_{\inf}^2\,\chi
+\dfrac{\epsilon_\theta v_\varphi^4}{\LCP}\,\sin 2\theta
\;=\; 0\;,
\label{EoM-nophi-chi2}
\vphantom{\Bigg|}\\
& & \biggl(1-\dfrac{v_\varphi^2}{\LCP^2}\biggr)
\Bigl(\ddot{\theta}+3H\dot{\theta}\Bigr)
+\epsilon_\theta v_\varphi^2 \sin 2\theta
+\dfrac{\mu_{\inf}^2}{\LCP}\,\chi
\;=\; 0\;.
\label{EoM-nophi-theta2}
\vphantom{\Bigg|}
\end{eqnarray}
Assuming,
\begin{equation}
\dfrac{v_\varphi^2}{\LCP^2} \;\ll\; 1 \;,\qquad
\dfrac{\epsilon_\theta v_\varphi^4}{\LCP} \;\ll\; \mu_{\inf}^2 \dfrac{M_p}{\xi}\;,
\label{ConditionA}
\end{equation}
the equation of motion for the inflaton $\chi$ becomes,
\begin{equation}
\ddot{\chi} \,+\, \left( 3 H +  \Gamma \right)\!\dot{\chi} \,+\, \mu_{\inf}^2\chi \;\approx\; 0\;,
\label{InflatonEOM}
\end{equation}
where $\Gamma$ is a friction term introduced to represent the loss of energy of the inflaton to all the other
degrees of freedom in the model.
We assume,
\begin{equation}
3H \,\ll\,\mu_{\inf}\;,\quad\Gamma \,\ll\, \mu_{\inf}\;, 
\label{ConditionB}
\end{equation}
and that their time dependences are much slower than $\mu_{\inf}$,
cf. Eq.~\eqref{ReheatingHubbleRange}.
The approximate solution to the above equation is then,
\begin{equation}
\chi(t)
\;\approx\; \chi_i \left( \frac{t_i}{t}\right)
e^{-\Gamma(t-t_i)/2}\cos\bigl[ \mu_{\inf}(t-t_i)\bigr] 
\;=\; \chi_i \left[\frac{H(t)}{H_i}\right]
e^{-\Gamma(t-t_i)/2}\cos\bigl[ \mu_{\inf}(t-t_i)\bigr] 
\;,
\label{Ch3PhiSolutionT}
\end{equation}
where $t_i$ is the time at which the reheating epoch begins, $\chi_i=\chi(t_i)$, and $H_i=H(t_i)$.
This solution indicates that the motion of $\chi(t)$ is oscillatory, with an angular frequency $\mu_{\inf}$, and an  amplitude predominantly attenuated  by Hubble damping early in reheating. 
Eqs.~\eqref{ConditionB} and \eqref{Ch3PhiSolutionT} imply
\begin{equation}
\dot{\chi}(t)
\;\approx\; 
-\mu_{\inf}\,\chi_i
 \left(\frac{t_i}{t}\right)
e^{-\Gamma(t-t_i)/2}\sin\bigl[ \mu_{\inf}(t-t_i)\bigr] 
\;,
\end{equation}
that is, $\dot{\chi}\sim \mu_{\inf}\chi$.

%
\subsubsection*{Driven Motion and Phase-Locked States}
\label{DMandPLS}

To derive the conditions under which a non-zero lepton number density $n_L = - 2v_\varphi^2 \dot{\theta}$ will be generated,
the dynamics of $\theta$ must be analysed within the background of reheating.
The equation of motion of $\theta$, taking into account Eq.~\eqref{ConditionA}, is
\begin{equation}
\ddot{\theta}
+\left(3H+\Gamma_\theta\right) \dot{\theta}
+\epsilon_\theta v_\varphi^2 \sin 2\theta
+\dfrac{1}{\LCP}\mu_{\inf}^2\,\chi
\;=\; 0\;,
\label{EoM-nophi-theta3}
\end{equation}
where we have introduced the friction term $\Gamma_{\theta}$ 
to account for $\theta$'s loss of energy to SM fields via the interactions of Eq.~\eqref{nu_int}.
Substituting Eq.~\eqref{Ch3PhiSolutionT}, we obtain
\begin{equation}
\ddot{\theta} 
\,+\, f(t)\,\dot{\theta}
\,+\, p\sin(2\theta)
\,+\, q(t)\cos\bigl[ \mu_{\inf}(t-t_i)\bigr]
\;=\; 0\;,
\label{Ch3thetaeom}
\end{equation}
where
\begin{equation}
f(t)\,=\, 3H(t) + \Gamma_{\theta} =(2/t) + \Gamma_{\theta}\;,\qquad
p\,=\,\epsilon_{\theta}v_\varphi^2\;,\qquad  
q(t)\,=\,\dfrac{\mu_{\inf}^2\chi_i}{\LCP}\frac{H(t)}{H_i}\;.
\end{equation}
This equation is analogous to that of a forced pendulum.
The term proportional to $\sin(2\theta)$ can be viewed as the gravitational force on the pendulum when it is at an angle $2\theta$ from vertical down.
$q(t)\cos[\mu_{\inf}(t-t_i)]$ is the external pushing force, and $f(t)$ the net friction.  
The added complexity in our case is that the amplitude of the external force $q(t)$ and the friction $f(t)$ on the pendulum both depend on $t$. 
However, the time evolution of $H(t)$, and consequently those of $f(t)$ and $q(t)$, is expected to be slow relative to the frequency of the driving force $\mu_{\inf}$ toward the end of the reheating epoch, cf. Eq.~\eqref{ReheatingHubbleRange}.
Therefore, to analyse the dynamics of $\theta$ during multiple oscillations of the inflaton $\chi$ within that time frame,
it suffices to replace $H(t)$ with a constant $H_d = H(t_d)$, where $t_d$ is the time at which driven motion occurs.
%

The equation of motion of $\theta$ now reads,
\begin{equation}
\ddot{\theta} 
\,+\, f_d\,\dot{\theta}
\,+\, p\sin(2\theta)
\,+\, q_d\cos\bigl[ \mu_{\inf}(t-t_i)\bigr]
\;=\; 0\;,
\label{Ch3thetaeom2}
\end{equation}
where $f_d=f(t_d$), $q_d=q(t_d)$.
This equation has been studied in a variety of contexts.
The solutions relevant in our scenario are 
those that increase or decrease monotonously in time with only small amplitude modulations,
\textit{i.e.} those of the form: 
\begin{equation}
\theta(t) \;=\; \theta_0 + \dfrac{n}{2}\mu_{\inf}(t-t_i) 
- \sum_{k=1}^{\infty}\alpha_k\sin\Bigl[
k\mu_{\inf}(t-t_i) + \delta_k
\Bigr]\;,\qquad
n\in\mathbb{Z}\;.
\label{PhaseLocked}
\end{equation}
Such solutions are known as \textit{phase-locked states} and are found in the study of the 
chaotic behaviour of the forced pendulum.
For instance, the conditions for phase-locked states to exist were investigated in the study of 
the chaotic behaviour of an electric current passing through a Josephson junction \cite{Pedersen:1980}.
There, it is shown that we require $p\simeq q_d = q(t_d)$.
This makes sense from the Leptogenesis point of view since 
for successful lepton-asymmetry generation, 
both the $L$ breaking $p$-term, and the $\mathcal{C}$ and $\mathcal{CP}$ violating $q(t)$-term
should contribute to the time evolution of $\theta$. 
Thus, during reheating we must achieve $p\simeq q(t_d)$, which we shall call the Sweet Spot Condition (SSC),
which determines the time $t_{d}$:
\begin{equation}
\underbrace{\vphantom{\bigg|}\;\epsilon_{\theta} v_\varphi^2\;}_{\displaystyle p} 
\simeq\; \underbrace{\frac{\mu_{\inf}^2\chi_i}{\LCP} \left( \frac{H_d}{H_i} \right)}_{\displaystyle q(t_d)}
\;.
\label{Ch3FinalTimeCondition2}
\end{equation}
When this condition is satisfied, 
rotational motion of the pendulum arises with an almost constant angular velocity $\dot{\theta}$.

For the phase-locked state solution, Eq.~\eqref{PhaseLocked}, the lepton number density $n_L$ is
calculated from the time average of $\dot{\theta}$ as ,
\begin{equation}
n_L 
\;=\; -2v_\varphi^2\underbrace{\langle\,\dot{\theta}\,\rangle}_{\displaystyle n\mu_{\inf}/2}
\;=\; -\left(\mu_{\inf} v_\varphi^2\right)n
\;.
\label{Ch3eq:barnodenEstimate}
\end{equation} 
Interestingly, this result depends on the integer $n$, 
where $n/2$ is the number of rotations of the phase $\theta$ 
per oscillation of the inflaton $\chi$.  
The value of $n$ is not uniquely determinable from the SSC and must be obtained from numerical simulations. 
In our previous work \cite{Bamba:2016vjs,Bamba:2018bwl} we found via repeated numerical analyses that the integer $n$ 
was given approximately by,
\begin{equation}
n 
\;\approx\; \dfrac{2\epsilon_{\theta} v_\varphi^2}{\mu_{\inf}^2}
\;=\; \dfrac{m_\theta^2}{\mu_{\inf}^2}
\;.
\label{approxn}
\end{equation}
%
%
Since $n$ is an integer,
\begin{equation}
m_\theta^2 \;=\;
2\epsilon_{\theta}v_\varphi^2 \,\ge\,\mu_{\inf}^2\;,
\label{mtheta-vs-mu}
\end{equation}
which implies that for driven motion to occur, 
the effective mass $m_\theta = \sqrt{2\epsilon_{\theta}}\,v_\varphi$ of the $\theta$ field must be greater than the mass 
$\mu_{\inf}$ of the inflaton $\chi$.
Using this approximate value of $n$, we can proceed with calculating the lepton-number density generated by the directed motion in $\theta$, which yields,
\begin{equation}
|n_L| \;=\; \mu_{\inf} v_\varphi^2 \,n \;\approx\; \frac{2\epsilon_{\theta}  v_\varphi^4}{\mu_{\inf}}\;.
\label{Ch3eq:barnoden}
\end{equation}
This is diluted to,
\begin{equation}
|n_L| \quad\to\quad
|n_L|_{\mathrm{reh}} \;=\; |n_L| \left(\frac{a_d}{a_\mathrm{reh}}\right)^3~,
\end{equation}
due to the expansion of the universe from $t_d$ to the end of reheating.
The entropy density at the end of reheating is \cite{Husdal:2016haj},
\begin{equation}
s_{\mathrm{reh}} \;=\; \dfrac{2\pi^2}{45}g_{*} T_{\mathrm{reh}}^3 \;,
\end{equation}
and thus the asymmetry parameter is,
\begin{eqnarray}
\eta_L^\mathrm{reh}
\;=\; \frac{|n_{L}|_{\mathrm{reh}}}{s_{\mathrm{reh}}} 
\;=\; 
\biggl(\dfrac{90}{2\pi^2 g_{*}}\biggr)
\biggl(\frac{\epsilon_{\theta}v_\varphi^4}{\mu_{\inf}T_\mathrm{reh}^3}\biggr)
\biggl(\frac{a_d}{a_\mathrm{reh}}\biggr)^3
\;.
\end{eqnarray}
The dilution factor can be written as,
\begin{equation}
\biggl(\frac{a_d}{a_\mathrm{reh}}\biggr)^3 
=\; \dfrac{\rho_{\mathrm{rad,reh}}}{\rho_{\mathrm{rad},d}}
\;=\;
\dfrac{H_{\mathrm{reh}}^2}
      {H_d^2}
\;=\;
\biggl(\frac{\pi^2 g_{*}}{90}\biggr) 
\biggl(\frac{T_\mathrm{reh}^4}{M_p^2 H_d^2}\biggr)\;,
\label{DilutionFactor}
\end{equation}
which allows us to write,
\begin{equation}
\eta_L^\mathrm{reh}
\;=\; 
\dfrac{\epsilon_{\theta} v_\varphi^4 T_\mathrm{reh}}{2\mu_{\inf} H_d^2 M_p^2} \;.
\end{equation}
Taking into account sphaleron redistribution \cite{Klinkhamer:1984di,Kuzmin:1985mm}, the final asymmetry generated is,
\begin{equation}
\eta_B 
\;=\; \frac{28}{79}\,\eta_L^{\mathrm{reh}} 
\;\simeq\;
0.18\;\frac{\epsilon_{\theta} v_\varphi^4 T_\mathrm{reh}}{\mu_{\inf} H_d^2 M_p^2} 
\;.
\label{etaB}
\end{equation}
This can be simplified by utilising the SSC, Eq.~\eqref{Ch3FinalTimeCondition2}, which leads to,
\begin{equation}
H_d \;\approx\; 
\underbrace{\biggl(\dfrac{2\epsilon_\theta v_\varphi^2}{\mu_{\inf}^2}\biggr)}_{\displaystyle \approx n}
\dfrac{\LCP H_i}{2\chi_i}
\;\approx\; \dfrac{n\mu_{\inf}}{2\sqrt{6}M_p}\LCP
\;\approx\; (3n\times 10^{-6})\LCP
\;,
\label{HdSSC}
\end{equation}
%
and finally,
\begin{eqnarray}
\frac{\eta_B}{\eta_B^\mathrm{obs}} \;\simeq\; 
\dfrac{0.3}{\epsilon_\theta}
\left(\dfrac{T_\mathrm{reh}}{10^{12}\,\mathrm{GeV}}\right)
\left(\dfrac{10^{18}\,\mathrm{GeV}}{\LCP}\right)^2
\;\approx\;
\dfrac{10^3}{\epsilon_\theta\sqrt{\xi}}
\left(\dfrac{10^{18}\,\mathrm{GeV}}{\LCP}\right)^2
\;,
\label{etaBoveretaBobs}
\end{eqnarray}
where we have used \cite{Aghanim:2018eyx}
\begin{equation}
\eta_B^\mathrm{obs} \;=\; 8.5\times 10^{-11}
\;.
\label{etaBobs}
\end{equation}
Therefore, a large asymmetry can be generated when considering $\epsilon_{\theta}<1$, $T_\mathrm{reh}>10^{12}$~GeV, and
$\LCP < 10^{18}\,\mathrm{GeV}$.
As this is a very rough estimate relying on various approximations and idealized efficiency, 
an over-abundance of the baryon asymmetry is encouraging.

It is worth noting that the inverse proportionality of the final expression, Eq.~\eqref{etaBoveretaBobs}, to $\epsilon_\theta$ is somewhat counter-intuitive
This is due to both $n_L$ and $H_d$ being proportional to $\epsilon_\theta$, and the expansion of the Universe, Eq.~\eqref{DilutionFactor},
diluting the asymmetry $n_L$ generated at $t=t_d$.
Of course, $\epsilon_\theta$ cannot be made arbitrary small since one must satisfy Eq.~\eqref{mtheta-vs-mu}.
The $\LCP$ dependence only enters through the dilution factor via $H_d$. 
That is, $\LCP$ only determines when lepton-number generation happens and not the amount of lepton number generated.

\subsection{Summary of Scales Involved}
\label{SSI}
Let us summarize the constraints on various scales and couplings involved in the discussion so far.

We have the masses $\mu_{\inf}$, $m_\varphi = \sqrt{2\lambda_\phi} v_\varphi$,
$m_\theta = \sqrt{2\epsilon_\theta} v_\varphi$, and the CP violation scale $\LCP$.
The inflaton mass $\mu_{\inf}\approx 3\times 10^{13}\,\mathrm{GeV}$, cf. Eq.~\eqref{mu_from_CMB}, is set by the CMB data and cannot be floated.
This  fixes the ratio $\lambda_h/\xi^2 \approx 5\times 10^{-10}$, cf. Eqs.~\eqref{mu_Higgs_inf} and \eqref{lambda_over_xi2},
and the reheating temperature $T_{\mathrm{reh}} = (3\times 10^{15}\,\mathrm{GeV})/\sqrt{\xi}$, cf. Eq.~\eqref{ReheatingTemp}.

Eq.~\eqref{Ch3FinalTimeCondition2} demands $m_\theta \geqslant \mu_{\inf}$, while the 
assumption that $\varphi$ is fixed to $v_\varphi$ during Pendulum Leptogenesis demands
$m_\varphi >m_\theta$.
Thus the ordering of the masses is
\begin{eqnarray}
\mu_{\inf} \;\le\; m_\theta & < & m_\varphi \cr
& \downarrow & \cr
\sqrt{\epsilon_\theta} & < & \sqrt{\lambda_\phi} \cr
& \downarrow & \cr
\epsilon_\theta & \ll & \lambda_\phi
\label{MassOrdering1}
\end{eqnarray}
$\LCP$ is constrained from the requirement that the SSC must be satisfied during reheating:
\begin{equation}
\begin{array}{c}
\dfrac{\mu_{\inf}}{\sqrt{6}\xi} \;<\; (3n\times 10^{-6})\LCP \; <\; \dfrac{\mu_{\inf}}{\sqrt{6}} \\
\downarrow \\
\dfrac{(4\times 10^{18}\,\mathrm{GeV})}{n\,\xi} \;<\; \LCP \;<\; \dfrac{(4\times 10^{18}\,\mathrm{GeV})}{n}
\end{array}
\label{LambdaCP_Range}
\end{equation}
cf. Eqs.~\eqref{ReheatingHubbleRange} and \eqref{HdSSC}.
$v_\varphi$ and $\epsilon_\theta$ are constrained by Eq.~\eqref{ConditionA}:
\begin{equation}
\dfrac{v_\varphi^2}{\LCP^2} \;\ll\; 1 \;,\qquad
\dfrac{\epsilon_\theta v_\varphi^4}{\LCP} \;\ll\; \mu_{\inf}^2\dfrac{M_p}{\xi} \;=\; 10^{41}\,\mathrm{GeV}^3\;.
\end{equation}
A possible set of parameters which satisfies all these constraints is,
\begin{eqnarray}
\LCP & = & 3\times 10^{16}\,\mathrm{GeV}\;,\cr
v_\varphi & = & 3\times 10^{15}\,\mathrm{GeV}\;,\cr
\epsilon_\theta & = & 4\times 10^{-7}\;,\cr
\lambda_\phi & = & 0.005\;,\cr
\lambda_h & = & 0.13\;,\cr
\kappa & = & 0\;,\cr
\xi & = & 2\times 10^4\;,\cr
n & = & 1\;, 
\end{eqnarray}
which corresponds to
\begin{eqnarray}
m_\theta \,=\, \mu_{\inf} & = & 3\times 10^{13}\,\mathrm{GeV}\;,\cr
m_\varphi & = & 3\times 10^{14}\,\mathrm{GeV}\;,\cr
T_{\mathrm{reh}} & = & 2\times 10^{13}\,\mathrm{GeV}\;,
\end{eqnarray}
and
\begin{equation}
\dfrac{\eta_B}{\eta_B^{\mathrm{obs}}}
\;\approx\; 10^{10}\;.
\end{equation}
The value of $\lambda_h$ above is its tree-level value at low energies.
If the RGE running of $\lambda_h$ is considered, the value of $\xi$ would be smaller, 
and the reheating temperature higher.

Let us now turn to the question of vacuum stability to see if it can be resolved
within these constraints.

\section{Higgs Vacuum Stability}
\label{Stability}

\subsection{Renormalization Group Running of $\lambda_h$}

The current experimentally determined values of the Standard Model (SM) parameters 
strongly suggest that the Higgs vacuum is only metastable \cite{EliasMiro:2011aa,Degrassi:2012ry,Lebedev:2012sy,Salvio:2013rja,Branchina:2014usa,Bezrukov:2014ipa}. 
This can be seen from the renormalization group (RGE) running of the Higgs quartic coupling $\lambda_h$.

\begin{figure}
	\centering
	\includegraphics[width=0.6\textwidth]{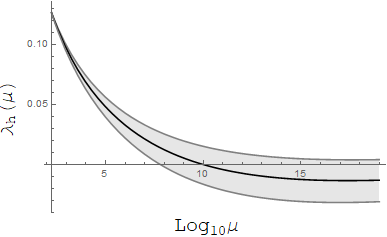}
	\caption{Running of the Higgs quartic coupling up to the Planck scale $M_p=2.4\times 10^{18}\,\mathrm{GeV}$. 
	The current experimental central values from Ref.~\cite{Tanabashi:2018oca} were used as input parameters with the the top quark pole mass, $m_t$, varied within its $3\sigma$ experimental constraints: 
	$m_t=173.1\pm 0.9$~GeV. The upper bound corresponds to $m_t=170.4$~GeV and the lower bound $m_t=175.8$~GeV.}
	 \label{higgsiggs}
\end{figure}

Figure~\ref{higgsiggs} shows the scale dependence of $\lambda_h^{\mathrm{SM}}(\mu)$,
the Higgs quartic coupling run with the 2-loop RGE's with only the SM particle content contributing, and 
with its low-energy value fixed to $\lambda_h^{\mathrm{SM}}(m_H) = 0.13$, cf. Eq.~\eqref{TreeLambdah}.
(See Appendix~\ref{2loopRGE} for the 2-loop RGE coefficients.)
The central solid line indicates $\lambda_h^{\mathrm{SM}}(\mu)$ when the Higgs and top masses are fixed to
their current central values of $m_H = 125.10\pm 0.14\,\mathrm{GeV}$ and $m_t = 173.1\pm 0.9\,\mathrm{GeV}$ (pole mass) \cite{Tanabashi:2018oca,Degrassi:2012ry}.
The upper and lower edges of the grey band depict the behaviour of $\lambda_h^{\mathrm{SM}}(\mu)$ when
the top mass is allowed to deviate from its central value by $-3\sigma$ (upper) and $+3\sigma$ (lower), respectively.
There is also some small dependence on the uncertainties in $m_H$ and $\alpha_s(M_Z)$ though these are not shown.
In Figure~\ref{higgsiggs}, for the central values of $m_H$ and $m_t$, 
$\lambda_h^{\mathrm{SM}}(\mu)$ goes negative around $\mu_{\mathrm{ins}} \approx 10^{10}\,\mathrm{GeV}$, 
and by the time it reaches the Planck scale, it is as negative as $\lambda_h^{\mathrm{SM}}(M_p)\simeq -0.013$.
Though it is still possible that the Higgs potential is stable up to the Planck scale if the
top mass is actually $3\sigma$ smaller than its central value, the likelihood of this is very small.

We reiterate here
that the stability of the Higgs vacuum is crucial for the Higgs inflation scenario. 
If the Higgs potential develops an instability at a high scale, it will be unable to support the slow roll of the Higgs towards a potential minimum. 
If it is metastable, the Higgs may become trapped in the true vacuum, never to reach the electroweak vacuum state we observe today.

\subsection{Scalar Threshold Effect}

The stability of the Higgs vacuum can be recovered by contributions of new particles to the running of $\lambda_h$. 
In the Pendulum Leptogenesis scenario, which includes the newly introduced complex scalar lepton $\phi$, 
there is no symmetry preventing the introduction of the following coupling between the Higgs doublet $\Phi$ and $\phi$:
\begin{equation}
\kappa(\Phi^\dagger\Phi)(\phi^\dagger\phi) \quad\to\quad \dfrac{\kappa}{4}h^2\varphi^2\;.
\end{equation}
Indeed, we have included this coupling in our scalar potential, Eq.~\eqref{scalar-potential-V}, from the beginning.
This coupling will provide an extra contribution to the RGE of $\lambda_h$ coming from $\varphi$.

\begin{figure}
	\centering
	\includegraphics[width=0.6\textwidth]{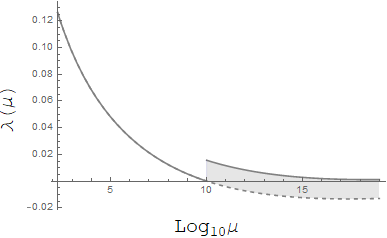}
	\caption{An example of the scalar threshold effect, in which $m_\varphi=10^{10}$~GeV, 
	$\lambda_\phi = 10^{-6}$, and $\kappa = 2.5\times 10^{-4}$. 
	In this case, Higgs vacuum stability is restored up to the Planck scale with the apparent discontinuity at the $ m_\varphi $ mass scale, near the instability scale, being due to the scalar threshold effect.  In the numerical calculations the 2-loop RGEs for the SM couplings are used, and are given in Appendix~\ref{2loopRGE} \cite{Salvio:2018rv}, with initial input values the central experimental values given in \cite{Tanabashi:2018oca}. }
	\label{scalar_thresh}
\end{figure}

The effect of the above coupling between the Higgs scalar $h$ and a SM-singlet real scalar $\varphi$ 
on the running of $\lambda_h(\mu)$ has  been explored previously \cite{Patt:2006fw,Lebedev:2011aq,Lebedev:2012zw,Ema:2017ckf,Salvio:2018rv}.
At scales above $m_\varphi$, the particle content of the model consists  of the SM$\;+\;\varphi$,
whereas below $m_\varphi$, it is  that of the SM with $\varphi$ integrated out.
Thus, the effective field theory that describes the system changes at $\mu = m_\varphi$, including
the running coupling $\lambda_h(\mu)$.
For $\mu < m_\varphi$ the running coupling is  that of the SM: $\lambda_h(\mu) = \lambda_h^{\mathrm{SM}}(\mu)$,
whereas for $\mu > m_\varphi$ it runs with contributions from $\varphi$.
At $\mu = m_\varphi$, the integrating out of the $\varphi$ degree of freedom leads to a finite shift \cite{EliasMiro:2012ay}:
\begin{equation}
\lambda_h^{\mathrm{SM}}(m_\varphi) \;=\;\lambda_h(m_\varphi)-\frac{\kappa^2(m_\varphi)}{4\lambda_\phi(m_\varphi)} \;.
\label{thresh}
\end{equation}
This scalar threshold effect in the evolution of $\lambda_h(\mu)$  at $\mu=m_\varphi$ is depicted in Figure~\ref{scalar_thresh}
for the case $m_\varphi = \mu_{\mathrm{ins}} = 10^{10}\,\mathrm{GeV}$,
$\lambda_\phi = 10^{-6}$, and $\kappa=2.5\times 10^{-4}$.
Note that regardless of the sign of $\kappa$, the shift term proportional to $\kappa^2/\lambda_\phi$ in Eq.~\eqref{thresh} implies a positive correction to $\lambda_h(\mu)$ at $\mu=m_\varphi$ when following its evolution from lower energies.
Furthermore, this shift can be significant even when both $\kappa$ and $\lambda_\phi$ are too small for the presence of $\varphi$ to have a non-negligible effect on the running of $\lambda_h(\mu)$ above $\mu=m_\varphi$.
In fact, 
as suggested by Figure~\ref{scalar_thresh}, it may be sufficiently large to maintain the positivity of
$\lambda_h(\mu)$ all the way up to the Planck scale.
Indeed, this can be accomplished by simply requiring,
\begin{equation}
\dfrac{\kappa^2(m_\varphi)}{4\lambda_\phi(m_\varphi)} \;>\; |\lambda_h^{\mathrm{SM}}(M_p)| \;\approx\; 0.01\;,
\label{kappa-bound1}
\end{equation}
or
\begin{equation}
\kappa^2(m_\varphi) \;\agt\; 0.04\,\lambda_\phi(m_\varphi)\;.
\label{kappa-bound2}
\end{equation}
Recall that the stability of the scalar potential, Eq.~\eqref{scalar-potential-V}, requires Eq.~\eqref{Vstability}:
\begin{equation}
4\lambda_h(m_\varphi)\lambda_\phi(m_\varphi) \;>\; \kappa^2(m_\varphi)\;,
\label{kappa-bound3}
\end{equation}
which implies that $\lambda_h^{\mathrm{SM}}(m_\varphi)$ of Eq.~\eqref{thresh} must be positive.
The positivity of $\lambda_h^{\mathrm{SM}}(m_\varphi)$ is a requirement on the scale $m_\varphi$, 
since this restricts $m_\varphi$ to be less than the
instability scale: $m_\varphi < \mu_{\mathrm{ins}}$.
This last requirement is incompatible with the order of scales that was assumed earlier, namely
$\mu_{\mathrm{ins}} < \mu_{\inf} \le m_\theta < m_\varphi$.

If we wish to simultaneously avoid the metastability of the Higgs vacuum, and  have successful Pendulum Leptogenesis, 
we must overcome the tension between these two scale orderings. 
For $\lambda_h(\mu)$ to remain positive up to the Planck scale, $m_\varphi$ must be less than or equal to the instability scale $\mu_{\mathrm{ins}}\simeq 10^{10}$~GeV, 
while Pendulum dynamics require $m_\theta$ to be near the scale of the inflaton $\mu_{\inf}\simeq 10^{13}$~GeV. 
Thus, we must re-evaluate the requirement $m_\theta < m_\varphi$ which was introduced to suppress the dynamics of
$\varphi$ during reheating.

In the following, we will perform a reanalysis of our model including  the full dynamics of $\varphi$
assuming $m_\varphi < \mu_{\mathrm{ins}}$.
We also set $\kappa\neq 0$ in order to have $\varphi$ contribute to the RG evolution of $\lambda_h$.
This mixing of $h$ and $\varphi$ means we must reanalysis the inflationary dynamics of our model with
possible contributions from both $h$ and $\varphi$ to the inflaton $\chi$.

\section{Dynamics of the Model Revisited}
\label{Revisit}

\subsection{Inflation: $\zeta\neq 0$ Case}
\label{Higgs_inf2}

We allow $\zeta\neq 0$ for the sake of generality.
For the purpose of Pendulum Leptogenesis, 
we wish to identify the Higgs scalar $h$ with the inflaton $\chi$.
The question is under what conditions can this identification be maintained.

We perform a conformal transformation from the Jordan frame to the Einstein frame
as in Eq.~\eqref{ConformalTransform} but this time with
\begin{equation}
\Omega^2 \;=\; 1 + \dfrac{\xi h^2}{M_p^2} + \dfrac{\zeta\varphi^2}{M_p^2}\;.
\end{equation}
In the inflationary regime where $\Omega^2 \gg 1$, the relevant terms in the action are
\begin{eqnarray}
S_E  & = & \int d^4x \,\sqrt{-\tilde{g}}\,
\Biggl[\,-\frac{M_p^2}{2}\,\tilde{R}
\,+\,\frac{\tilde{g}^{\mu\nu}}{2}
M_p^2\dfrac{3(\partial_\mu\Omega^2)(\partial_\nu\Omega^2)}{2\Omega^4} 
\,-\,\frac{1}{\Omega^4}\,V(h,\varphi,\theta)
\,+\,\cdots
\,\Biggr]
\;,\qquad
\end{eqnarray}
where the scalar potential $V(h,\varphi,\theta)$ is that given in Eq.~\eqref{scalar-potential-V}.
The inflaton field $\chi$ can be identified, as in Eq.~\eqref{chi-inflation}, with,
\begin{equation}
\chi\;\approx\; \sqrt{\dfrac{3}{2}}M_p\ln\Omega^2
\;=\; \sqrt{\dfrac{3}{2}}M_p\ln\left(1+\dfrac{\xi h^2}{M_p^2}+\dfrac{\zeta\varphi^2}{M_p^2}\right)\;.
\label{chi-inflation2}
\end{equation}
The direction of $\chi$ in the 2D $h$-$\varphi$ space is determined by $V(h,\varphi,0)/\Omega^4$,
where we set $\theta=0$ to suppress the $L$-violating term.
Let
\begin{equation}
r^2 \;=\; \xi h^2 + \zeta\varphi^2\;,\quad
\tan\delta \;=\; \dfrac{\sqrt{\zeta}\varphi}{\sqrt{\xi}h}
\quad\to\quad
\sqrt{\xi}h \;=\; r\cos\delta\;,\quad
\sqrt{\zeta}\varphi \;=\; r\sin\delta\;.
\end{equation}
Then $\Omega^2 = 1 + (r^2/M_p^2)$ and,
\begin{equation}
\dfrac{1}{\Omega^4}V(h,\varphi,0)
\quad\xrightarrow{r\gg M_p}\quad
\dfrac{M_p^4}{4}
\biggl[\;
\underbrace{
 \underbrace{\left(\dfrac{\lambda_h}{\xi^2}\right)}_{\displaystyle A}c_\delta^4 
+\underbrace{\left(\dfrac{\kappa}{\xi\zeta}\right)}_{\displaystyle B}c_\delta^2 s_\delta^2
+\underbrace{\left(\dfrac{\lambda_\phi}{\zeta^2}\right)}_{\displaystyle C}s_\delta^4
}_{\displaystyle \lambda_\mathrm{eff}(s_\delta^2)}
\;\biggr]
\left(1-\dfrac{1}{\Omega^2}\right)^2
\;,
\end{equation}
where $c_\delta =\cos\delta$, $s_\delta =\sin\delta$.
The angle $\delta$ minimizes
\begin{eqnarray}
\lambda_\mathrm{eff}(s_\delta^2) 
& = & A c_\delta^4 + B c_\delta^2 s_\delta^2 + C s_\delta^4
\;,
\end{eqnarray}
that is, the inflaton $\chi$ evolves along the deepest valley of $V(h,\varphi,0)/\Omega^4$.
Note that $s_\delta^2$ is resricted to the range $0\le s_\delta^2 \le 1$.
The stability of the potential demands Eq.~\eqref{Vstability},
\begin{equation}
4\lambda_h\lambda_\phi \,>\, \kappa^2 
\quad\to\quad 4AC \,>\, B^2
\quad\to\quad -2\sqrt{AC}\,<\,B\,<\,2\sqrt{AC}
\;.
\label{Brestriction}
\end{equation}
Let us assume $0<A\ll C$,
\begin{equation}
\dfrac{\lambda_h}{\xi^2} \;\ll\; \dfrac{\lambda_\phi}{\zeta^2}\;,
\label{AllC}
\end{equation}
but allow $B = \kappa/(\xi\zeta)$ to have either sign.
The $\zeta=0$ case is included in this restriction.
As shown in Appendix~\ref{AppLambdaEffMin}, 
if $2A \le B < 2\sqrt{AC}$,
\begin{equation}
\dfrac{2\zeta\lambda_h}{\xi} \;\le\; \kappa \;<\; 2\sqrt{\lambda_h\lambda_\phi} \;,
\label{kappaRange1}
\end{equation}
then the minimum of $\lambda_\mathrm{eff}(s_\delta^2)$ is at $s_\delta^2=0$ where it is equal to,
\begin{equation}
\lambda_\mathrm{eff}(0) \;=\; A \;=\; \dfrac{\lambda_h}{\xi^2}\;.
\end{equation}
In this case, the inflaton $\chi$ consists purely of the Higgs scalar $h$.
Note that due to Eq.~\eqref{AllC}, $\kappa$ may not need to be significantly fine-tuned to fall into this range.

If $-2\sqrt{AC} < B < 2A$,
\begin{equation}
-2\sqrt{\lambda_h\lambda_\phi} < \kappa < \dfrac{2\zeta\lambda_h}{\xi}\;,
\label{kappaRange2}
\end{equation}
then the minimum of $\lambda_\mathrm{eff}(s_\delta^2)$ is at,
\begin{equation}
s_\delta^2 \;=\; \breve{s}_\delta^2
\;\equiv\; \dfrac{A-(B/2)}{A-B+C} 
\;<\; \dfrac{\sqrt{A}}{\sqrt{A}+\sqrt{C}}
\;\ll\; 1\;,
\label{breve-s2}
\end{equation}
where it has the value,
\begin{equation}
\lambda_{\mathrm{eff}}(\breve{s}_\delta^2)
\;=\; \dfrac{AC-(B^2/4)}{A-B+C}
\;<\; A
\;,
\end{equation}
which is positive, as it should be, due to Eq.~\eqref{Brestriction}.
In this case, the inflaton $\chi$ is a mixture of $h$ and $\varphi$, but the condition $A\ll C$,
cf. Eq.~\eqref{AllC}, ensures Eq.~\eqref{breve-s2},
$\breve{s}_\delta^2 \ll 1$, 
so that the admixture of $\varphi$ in $\chi$ is very small.
In the limit $\zeta\to 0$, we have $\breve{s}_{\delta}^2\to 0$, and 
$\lambda_{\mathrm{eff}}(\breve{s}_\delta^2)\to (\lambda_h - \kappa^2/4\lambda_\phi)/\xi^2$.

Therefore, by requiring Eqs.~\eqref{AllC} and \eqref{kappaRange1} or \eqref{kappaRange2},
the inflaton $\chi$ can be made to consist only (in the case of Eq.~\eqref{kappaRange1})
or mostly (in the case of Eq.~\eqref{kappaRange2}) of the Higgs scalar $h$.
In both cases, the relation between $\chi$ and $h$ can be considered to be given by
Eq.~\eqref{chi-h-relation}, and the inflaton potential will be given by, 
\begin{equation}
V(\chi) 
\;=\; \dfrac{V(h,\varphi,0)}{\Omega^4}
\;\approx\; \dfrac{\breve{\lambda}_\mathrm{eff}M_p^4}{4}\left(1-\dfrac{1}{\Omega^2}\right)^2
\;=\; \dfrac{3}{4}\mu_{\inf}^2 M_p^2
\left[ 1 - e^{-\sqrt{\frac{2}{3}}(\chi/M_p)} \right]^2
\;,
\end{equation}
where $\breve{\lambda}_{\mathrm{eff}}$ is the minimum value of $\lambda_{\mathrm{eff}}(s_\delta^2)$
and,
\begin{equation}
\mu_{\inf}^2 \;=\; \dfrac{\breve{\lambda}_{\mathrm{eff}}M_p^2}{3}\;.
\label{mu_Higgs_inf2}
\end{equation}
This expression encompasses Eq.~\eqref{mu_Higgs_inf} for the $\zeta=0$, $\kappa=0$ case
in which $\breve{\lambda}_{\mathrm{eff}}=\lambda_h/\xi^2$.

\bigskip
\subsection{Reheating Dynamics and Pendulum Leptogenesis Revisited}
\label{Reheating2}

\subsubsection*{The Effective Action}

During the reheating epoch, $\Omega^2 \gtrsim 1$,  
and all the terms in the action that were neglected during inflation due to $\Omega^2 \gg 1$ must now be considered.
The relevant terms are now,
\begin{eqnarray}
S_E 
& = & \int d^4x \,\sqrt{-\tilde{g}}\,
\Bigg[\,-\,\dfrac{M_p^2}{2}
\left\{\tilde{R}
- \tilde{g}^{\mu\nu}\dfrac{3(\partial_\mu\Omega^2)(\partial_\nu\Omega^2)}{2\Omega^4}
\right\}
\cr
& & \qquad\qquad\qquad
+\;\dfrac{\tilde{g}^{\mu\nu}}{2\Omega^2}
\Bigl(\partial_\mu h\,\partial_\nu h 
+ \partial_\mu \varphi\,\partial_\nu \varphi 
+ \varphi^2 \partial_\mu \theta\,\partial_\nu \theta
\Bigr)
\,-\,\dfrac{1}{\Omega^4}V(h,\varphi,\theta)
\,+\,\dfrac{1}{\Omega^4}\mathcal{L}_{\cancel{\mathcal{CP}}}
\Biggr]~.
\cr
& &
\label{higgs-phi_E}
\end{eqnarray}
As discussed in the previous subsection, when Eq.~\eqref{kappaRange1} is satisfied, the
inflaton $\chi$ consists of only the Higgs scalar $h$, and when Eq.~\eqref{kappaRange2} is satisfied the
inflaton $\chi$ still consists mostly of the Higgs scalar $h$ with only a very small admixture of $\varphi$.
Under this assumption, we replace the kinetic terms of $\Omega^2$ and $h$ in the above action 
with that of $\chi$ as in Eq.~\eqref{chi-def} such that,
\begin{eqnarray}
S_E 
& \approx & \int d^4x \,\sqrt{-\tilde{g}}\,
\Bigg[\,-\,\dfrac{M_p^2}{2}\tilde{R}
+ \dfrac{\tilde{g}^{\mu\nu}}{2}
\left\{
  \partial_\mu\chi\,\partial_\nu\chi 
+ \partial_\mu \varphi\,\partial_\nu \varphi 
+ \varphi^2 \partial_\mu \theta\,\partial_\nu \theta
\right\}
\cr
& & \qquad\qquad\qquad
\,-\,V(\chi,\varphi,\theta)
\,-\,\dfrac{\tilde{g}^{\mu\nu}}{\LCP}\,\varphi^2
(\partial_\mu\theta)(\partial_\nu\chi)
\Biggr]
\;.
\end{eqnarray}
Here, $V(\chi,\varphi,\theta)$ is the scalar potential $V(h,\varphi,\theta)$ in which
the Higgs scalar $h$ has been replaced by the inflaton $\chi$ using Eq.~\eqref{chi-h-relation}:
\begin{eqnarray}
V(h,\varphi,\theta)
& = &
 \dfrac{\lambda_h}{4}(v_h^2-h^2)^2 
+\dfrac{\kappa}{4}(v_h^2-h^2)(v_\varphi^2-\varphi^2)
+\dfrac{\lambda_\phi}{4}(v_\varphi^2-\varphi^2)^2
+\epsilon_\theta\varphi^4\sin^2\theta
\vphantom{\Bigg|}\cr
& \downarrow & \vphantom{\Big|}\cr
V(\chi,\varphi,\theta)
& = & 
 \dfrac{\lambda_h}{4}\biggl(v_h^2-\sqrt{\dfrac{2}{3}}\dfrac{M_p}{\xi}\,\chi\biggr)^2 
+\dfrac{\kappa}{4}\biggl(v_h^2-\sqrt{\dfrac{2}{3}}\dfrac{M_p}{\xi}\,\chi\biggr) (v_\varphi^2-\varphi^2)
+\dfrac{\lambda_\phi}{4}(v_\varphi^2-\varphi^2)^2
+\epsilon_\theta\varphi^4\sin^2\theta
\vphantom{\Bigg|}\cr
& \approx &
\dfrac{1}{2}\mu_{\inf}^2\,\chi^2
 -K\chi
 (v_\varphi^2 - \varphi^2)
+\dfrac{\lambda_\phi}{4}(v_\varphi^2-\varphi^2)^2
+\epsilon_\theta\varphi^4\sin^2\theta
\;,
\vphantom{\Bigg|}
\end{eqnarray}
where
\begin{equation}
K \;\equiv\; \dfrac{\kappa M_p}{2\sqrt{6}\xi}\;.
\end{equation}
Note that a $\chi\varphi^2$ coupling term exists when $\kappa\neq 0$.
In this form, the potential minima are at,
\begin{equation}
\vev{\chi} \;=\; 0\;,\qquad
\vev{\varphi^2} \;=\; v_\varphi^2\;,\qquad
\vev{\theta} \;=\; n\pi\;,\quad
n\in\mathbb{Z}\;.
\end{equation}
The stability of these points require,
\begin{equation}
\lambda_\phi \mu_{\inf}^2 \; > \; 2K^2\;,
\end{equation}
which is guaranteed by Eq.~\eqref{Vstability}.

The main difference between our original analysis and the current one is the dynamical nature of $\varphi$ 
and the motion induced by its direct coupling to the inflaton: $\chi\varphi^2$. 
To obtain directed motion in $\theta$, as in the original scenario, $\varphi$ must become trapped around its minimum with only small oscillations around it sometime during the reheating epoch, with sufficient time left for the $\dot{\chi}\dot{\theta}$ interaction to do its job.

\subsubsection*{Imposition of Various Viability Conditions}

Selecting the flat Friedmann-Robertson-Walker metric with scale factor $a(t)$ and ignoring spatial dependence as before, 
the equations of motion obtained by varying $\varphi$, $\chi$ and $\theta$ are given respectively by
\begin{eqnarray}
& & \ddot{\varphi} +3H\dot{\varphi} 
    -\left(\dot{\theta}^2-\frac{2\dot{\chi}\dot{\theta}}{\LCP}
     \right)
\varphi 
+\frac{\partial V}{\partial \varphi}
\;=\;0\;,
\vphantom{\Bigg|}
\label{EoM_phi} 
\\
& & \ddot{\chi} +3H\dot{\chi} 
-\frac{\varphi^2}{\LCP}
 \left[
    \ddot{\theta}
    +\left(3H +\frac{2\dot{\varphi}}{\varphi}
     \right)\dot{\theta}
 \right]+\frac{\partial V}{\partial \chi}
\;=\;0\;,
\vphantom{\Bigg|}
\label{EoM_h}
\\
& & \ddot{\theta} 
+\left(3H +\frac{2\dot{\varphi}}{\varphi}
 \right)\dot{\theta} 
-\frac{1}{\LCP}
 \left[
    \ddot{\chi}
    +\left(3H +\frac{2\dot{\varphi}}{\varphi}
     \right)\dot{\chi}
 \right] 
+\frac{1}{\varphi^2}\frac{\partial V}{\partial \theta}
\;=\;0\;.
\vphantom{\Bigg|}
\label{EoM_theta}
\end{eqnarray}
Compare with Eqs.~\eqref{EoM-nophi-chi1} and \eqref{EoM-nophi-theta1}.
Keeping in mind Eq.~\eqref{AllC}, which is already imposed in order to have $\chi\sim h^2$, 
and  Eq.~\eqref{kappaRange1} for inflation driven purely by $h$,
we wish to find the conditions under which the following requirements can be satisfied:
\begin{enumerate}
\item The oscillatory dynamics of the inflaton $\chi$ is little affected by those of $\varphi$, $\theta$, or any of the SM fields.
\item $\varphi$ becomes trapped close to its vacuum expectation value $v_\varphi$ not too far into the reheating epoch
so that there is ample time to generate the rotational motion of $\theta$.
\item The SSC, Eq.~\eqref{Ch3FinalTimeCondition2}, can be satisfied after $\varphi$ is trapped.
\item The Higgs stability conditions, $m_\varphi = \sqrt{2\lambda_\phi}v_\varphi < \mu_{\mathrm{ins}}$ and Eq.~\eqref{kappa-bound2}, are satisfied.
\end{enumerate}
Previously, the effects of $\varphi$, $\theta$, and the SM fields on the dynamics of the inflaton $\chi$
were all absorbed into a constant friction term $\Gamma$, cf. Eq.~\eqref{InflatonEOM}.
The conditions 2 and 3 were satisfied by suppressing the dynamics of $\varphi$ by choosing $m_\phi \gg m_\theta$.
Here, we refrain from that option for the sake of 4.
Let us look at these requirements one by one.

\subsubsection*{Dynamics of $\chi$ and $\theta$}

We first look at the dynamics of $\chi$ and $\theta$ to clarify what requirements need to be placed on the
dynamics of $\varphi$ for Pendulum Leptogenesis to proceed as before.

Rearranging Eqs.~\eqref{EoM_h} and \eqref{EoM_theta}, writing out the force terms explicitly, 
and introducing friction terms to account for the interaction of $\chi$ and $\theta$ with SM fields,
we obtain,
\begin{eqnarray}
& & 
\left(1-\frac{\varphi^2}{\LCP^2}
\right)
\Bigl(
   \ddot{\chi} +3H\dot{\chi}
\Bigr)
\,+\, 
\left\{
  \Gamma
  -
  \left(\dfrac{\varphi^2}{\LCP^2}\right)
  \left(\dfrac{2\dot{\varphi}}{\varphi}\right)
\right\}\dot{\chi}
\vphantom{\Bigg|}\cr
& & \qquad\qquad\qquad\qquad\quad
\,+\,\mu_{\inf}^2\chi
\,+\,K(\varphi^2 - v_\varphi^2)
\,+\,\frac{\epsilon_{\theta}}{\LCP}\;\varphi^4\sin 2\theta
\;=\; 0\;,
\vphantom{\Bigg|}
\label{EoM_chi1}
\\
& &
\left(1-\frac{\varphi^2}{\LCP^2}
\right)
\left\{
   \left(
      \ddot{\theta} 
      +3H\dot{\theta}
   \right)
  +\left(
      \dfrac{2\dot{\varphi}}{\varphi}
   \right) \dot{\theta}
\right\}
\,+\,\Gamma_\theta\,\dot{\theta} 
\,+\,\epsilon_{\theta}\,\varphi^2\sin 2\theta
\vphantom{\Bigg|}\cr
& & \qquad\qquad\qquad\qquad
\,+\,\dfrac{1}{\LCP}
\left\{
 \mu_{\inf}^2\,\chi
\,+\,K(\varphi^2 - v_\varphi^2)
\,-\left(\dfrac{2\dot{\varphi}}{\varphi}\right)\dot{\chi}
\right\}
\;=\; 0\;.
\label{EoM_theta1}
\vphantom{\Bigg|}
\end{eqnarray}
Compare with Eqs.~\eqref{EoM-nophi-chi2} and \eqref{EoM-nophi-theta2}.
As before,
we would like the oscillatory dynamics of the inflaton $\chi$ to be little affected by the dynamics of $\varphi$ or $\theta$.
That is, we wish to find the conditions under which the Eq.~\eqref{EoM_chi1} can be well approximated by
Eq.~\eqref{InflatonEOM}, and determine how Eq.~\eqref{EoM-nophi-theta3} will be modified.

We first require,
\begin{equation}
\dfrac{\varphi^2}{\LCP^2} 
\;\ll\; 1 \;,\qquad
\dfrac{\epsilon_\theta \varphi^4}{\LCP} 
\;\ll\; \mu_{\inf}^2\dfrac{M_p}{\xi}\;,
\label{ConditionAprime}
\end{equation}
which would be the same condition as Eq.~\eqref{ConditionA}
if $\varphi = O(v_\varphi)$.
The equations become,
\begin{eqnarray}
& &
\Bigl(
   \ddot{\chi} +3H\dot{\chi}
\Bigr)
\,+\,\Gamma\,\dot{\chi}
\,+\,\mu_{\inf}^2\chi
\,+\,K(\varphi^2 - v_\varphi^2)
\;=\; 0\;,
\vphantom{\Bigg|}
\label{EoM_chi2}
\\
& &
\left(
   \ddot{\theta} 
   +3H\dot{\theta}
\right)
+\left\{
   \Gamma_\theta
   +\left(
       \dfrac{2\dot{\varphi}}{\varphi}
    \right)
 \right\} \dot{\theta} 
\,+\,\epsilon_{\theta}\,\varphi^2\sin 2\theta
\vphantom{\Bigg|}\cr
& & \qquad\qquad\qquad\qquad
\,+\,\dfrac{1}{\LCP}
\left\{
 \mu_{\inf}^2\,\chi
\,+\,K(\varphi^2 - v_\varphi^2)
\,-\left(\dfrac{2\dot{\varphi}}{\varphi}\right)\dot{\chi}
\right\}
\;=\; 0\;.
\label{EoM_theta2}
\vphantom{\Bigg|}
\end{eqnarray}
%
%
%
In order for $\theta$ to undergo a well-defined directed rotational motion, $\varphi$ should be trapped in one of its
potential wells, e.g. at $\varphi = v_\varphi$, and only undergo small oscillations around it, at least toward the
latter half of the reheating epoch.
Since the motion of $\varphi$ will be driven by the oscillation of $\chi$, we expect it to oscillate  with
the same frequency $\mu_{\inf}$, that is,
\begin{eqnarray}
\varphi & \sim & v_\varphi + \alpha \cos\left(\mu_{\inf}t+\delta_\varphi\right)
\cr
& \downarrow & \cr
\dfrac{2\dot{\varphi}}{\varphi} & \sim & \dfrac{2\mu_{\inf}\alpha}{v_\varphi} \;,\vphantom{\Bigg|}\cr
\varphi^2 - v_\varphi^2 & \sim & 2v_\varphi\alpha \cos\left(\mu_{\inf}t+\delta_\varphi\right) + O(\alpha^2)\;.\vphantom{\Bigg|}
\end{eqnarray}
We will see later that
\begin{equation}
\dfrac{2\alpha}{v_\varphi} \;\sim\; \mathcal{O}(1)\;,
\label{ConditionC}
\end{equation}
\begin{equation}
\dfrac{2\dot{\varphi}}{\varphi}
\;\sim\; \mu_{\inf}
\;.
\end{equation}
Let us demand,
\begin{eqnarray}
\left| K(\varphi^2 - v_\varphi^2) \right| 
& \ll &  
\left| \mu_{\inf}^2\chi \right| 
\cr
& \downarrow & \cr
\left| 2Kv_\varphi\alpha \right| 
& \ll & 
\left| \mu_{\inf}^2 \dfrac{M_p}{\xi} \right|
\cr
& \downarrow & \cr
\dfrac{|\kappa|v_\varphi^2}{2\sqrt{6}\mu_{\inf}^2}\dfrac{2\alpha}{v_\varphi} 
& \ll & 1
\cr
& \downarrow & \cr
\dfrac{|\kappa|v_\varphi^2}{2\sqrt{6}\mu_{\inf}^2}
& \ll & 1\;.
\label{ConditionK}
\end{eqnarray}
The equations become,
\begin{eqnarray}
& & 
\Bigl(
   \ddot{\chi} +3H\dot{\chi}
\Bigr)
\,+\, 
\Gamma\,\dot{\chi}
\,+\,\mu_{\inf}^2\chi
\;=\; 0\;,
\vphantom{\Bigg|}
\label{EoM_chi3}
\\
& &
\left(
   \ddot{\theta} 
   +3H\dot{\theta}
\right)
+\left\{
   \Gamma_\theta
   +\left(
       \dfrac{2\dot{\varphi}}{\varphi}
    \right)
 \right\} \dot{\theta} 
\,+\,\epsilon_{\theta}\,\varphi^2\sin 2\theta
\vphantom{\Bigg|}\cr
& & \qquad\qquad\qquad\qquad
\,+\,\dfrac{1}{\LCP}
\left\{
 \mu_{\inf}^2\,\chi
\,-\left(\dfrac{2\dot{\varphi}}{\varphi}\right)\dot{\chi}
\right\}
\;=\; 0\;.
\label{EoM_theta3}
\vphantom{\Bigg|}
\end{eqnarray}
The equation for the inflaton $\chi$ is now in its desired form.
The $(2\dot{\varphi}/\varphi)$ terms that appear in the equation of motion for $\theta$ are non-negligible.
$(2\dot{\varphi}/\varphi)$ in the coefficient of $\dot{\theta}$ dominates over $(3H+\Gamma_\theta)$ and renders the friction time-dependent and oscillatory. 
Since $\dot{\chi}\sim \mu_{\inf}\chi$, we have,
\begin{equation}
\left(\dfrac{2\dot{\varphi}}{\varphi}\right)\dot{\chi}
\;\sim\; \left(\dfrac{2\alpha}{v_\varphi}\right) \mu_{\inf}^2 \chi
\;\sim\; \mu_{\inf}^2 \chi
\;.
\end{equation}
Therefore, this term is of the same order as the $\mu_{\inf}^2\chi$ term.
Both terms will contribute to driving the motion of $\theta$,
though the time-dependence of the friction will modify the dynamics from
the original setup.
We perform a thorough analysis of its effect in the next subsection.
It is demonstrated that phase-locked states with $\dot{\theta}\sim\mu_{\inf}$ are generated as before within this framework.

\subsubsection*{Dynamics of $\varphi$}

Let us consider the dynamics of $\varphi$.
As we found above, we need the motion of $\varphi$ to be limited to small oscillations around
$\varphi = v_\varphi$ (or $\varphi = -v_\varphi$)
for rotational motion of $\theta$ in one direction to be generated.

Eq.~\eqref{EoM_phi} with the force terms written out explicitly 
and an added friction term is
\begin{equation}
\Bigl\{
 \ddot{\varphi} 
+\bigl(3H + \Gamma_\varphi\bigr)\dot{\varphi} 
\Bigr\}
+\left\{
 -\dfrac{m_\varphi^2}{2}
 -\dot{\theta}^2
 +2\biggl(\!
    -K \chi
    +\frac{1}{\LCP}\,
     \dot{\theta}\dot{\chi}
   \biggr) 
 \!\right\}\varphi
+\Bigl(
    \lambda_\phi 
    +4\epsilon_{\theta} \sin^2\theta
 \Bigr)\varphi^3
\;=\; 0
\;.
\vphantom{\Bigg|}
\label{EoM_phi1}
\end{equation}
The friction term $\Gamma_\varphi$ comes from the decay of $\varphi$ to right-handed neutrinos through
the interaction shown in Eq.~\eqref{nu_int}.  

The $m_\varphi^2\varphi/2$ and $\lambda_\phi \varphi^3$ terms compete with each other to push $\varphi$ toward 
the potential minima at $\varphi = \pm v_\varphi = \pm \sqrt{m_\varphi^2/2\lambda_\phi}$; the $m_\varphi^2\varphi/2$ term forces
$\varphi$ away from $\varphi = 0$, and the $\lambda_\phi \varphi^3$ term pushes $\varphi$ toward $\varphi=0$.

\begin{figure}
	\centering
	\includegraphics[width=0.6\textwidth]{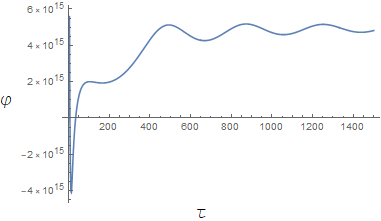}
	\caption{Plot of the oscillation of $\varphi$ during reheating, showing its evolution and subsequent trapping around one of its potential minima, as governed by Eq.~\eqref{EoM_phi1} when neglecting the production of driven motion in $\theta$, 
that is $\dot{\theta} \simeq \dot{\chi}/\LCP$. 
	The horizontal axis is in the units $\tau=\mu_{\inf} t$, and the example parameters used are  
	$\lambda_\phi = 10^{-8}$, $\epsilon_\theta=4 \times 10^{-5}$, $\xi=2\times 10^{4}$, $\zeta=100$, and 
	$v_{\varphi}=5\times 10^{15}$~GeV. }
	\label{varphi}
\end{figure}

The $\dot{\theta}^2$ term is the centrifugal barrier,
which also forces $\varphi$ away from $\varphi=0$ and contributes to trapping $\varphi$.
During reheating, the motion of $\theta$ is driven by the oscillation of $\chi$, so it will either oscillate with the
same frequency $\mu_{\inf}$ so that $\dot{\theta} \sim \mu_{\inf}\theta$, 
or enter into a phase locked state with $\dot{\theta} \sim n\mu_{\inf}/2$, cf. Eq.~\eqref{Ch3eq:barnodenEstimate}.
In either case, $\dot{\theta}^2$ scales as,
\begin{equation}
\dot{\theta}^2 \;\sim\; \mu_{\inf}^2\;,
\end{equation}
and if $m_\varphi < \mu_{\mathrm{ins}} \ll \mu_{\inf}$, the centrifugal barrier term can be expected to dominate over the $m_\varphi^2/2$ term throughout the reheating epoch.
On the other hand,
the $4\epsilon_\theta\sin^2\theta\,\varphi^3$ term, which contributes to pushing $\varphi$ toward zero, 
dominates over the $\lambda_\phi\varphi^3$ term
since $\lambda_\phi \ll \epsilon_\theta$ due to our assumption $m_\varphi < \mu_\mathrm{ins} \ll \mu_{\inf} \le m_\theta$,

The $K$ and the $\LCP$ terms couple $\varphi$ to the inflaton $\chi$.
These terms oscillate around zero and drive the motion of $\varphi$.
Since the oscillation amplitude of $\chi$ evolves from $M_p$ to $M_p/\xi$ during reheating, cf. Eq.~\eqref{Uchi}, the 
amplitude of the $K\chi$ term evolves as,
\begin{eqnarray}
|K\chi| 
\;\approx\; |K|M_p \to |K|\dfrac{M_p}{\xi} 
& = & \dfrac{|\kappa|M_p^2}{2\sqrt{6}\xi} \to \dfrac{|\kappa|M_p^2}{2\sqrt{6}\xi^2} \vphantom{\Bigg|}\cr
& \approx & \mu_{\inf}^2 \left(\dfrac{10^9|\kappa|}{\xi} \to \dfrac{10^9|\kappa|}{\xi^2}\right)\;. \vphantom{\Bigg|}
\end{eqnarray}
For the amplitude of other term, we use $\dot{\theta}\sim \mu_{\inf}$, $\dot{\chi}\sim \mu_{\inf}\chi$ and estimate,
\begin{equation}
\biggl|\dfrac{\dot{\theta}\dot{\chi}}{\LCP}\biggr| 
\;\approx\; \mu_{\inf}^2\left(\dfrac{M_p}{\LCP} \to \dfrac{M_p}{\LCP\xi}\right)
\;.
\end{equation}
Since $\xi\sim 10^4$, by judicious choices of small $\kappa$ and large $\LCP$, we can
suppress these terms in comparison to the centrifugal barrier $\sim\mu_{\inf}^2$, in particular, toward the end of the reheating epoch.
So let us require,
\begin{equation}
\dfrac{10^9|\kappa|}{\xi^2} \;\ll\; 1\;,\qquad
\dfrac{M_p}{\LCP\xi} \;\ll\; 1\;.
\label{ConditionD}
\end{equation}
Then, toward the latter half of the reheating period, the equation of motion of $\varphi$ is,
\begin{equation}
\ddot{\varphi} + \Bigl(3H+\Gamma_\varphi\Bigr)\dot{\varphi} 
-\biggl(\dfrac{m_\varphi^2}{2}+\dot{\theta}^2\biggr)\varphi 
+ \Bigl(\lambda_\phi + 4\epsilon_\theta\sin^2\theta \Bigr)\varphi^3 \;\approx\; 0\;.
\end{equation}
We keep both the $m_\varphi^2$ and $\lambda_\phi$ terms to prevent the linear and cubic
terms from periodically vanishing  when $\theta$ is oscillatory.
The motion of $\varphi$ is now determined by the competition between the push away from
$\varphi=0$ by the linear term, and the push toward $\varphi=0$ by the cubic term.
Both are driven indirectly by the inflaton $\chi$ via $\theta$.

This equation is quite promising for the entrapment of $\varphi$ into either $\varphi > 0$ or $\varphi <0$,
which is a prerequisite for $\theta$ entering into a phase-locked state.
Early in the reheating epoch, $\theta$ can be expected to be provided with sufficient energy from the inflaton
$\chi$ to jump over the peaks of the potential $4\epsilon_\theta\sin^2\theta$ with ease.
Its oscillation amplitude will be very large, covering multiple potential wells per oscillation.
In that situation, $\dot{\theta}$ will also be very large, oscillating with
amplitude $\mu_{\inf}\mathcal{A}$, where $\mathcal{A}$ is the amplitude of $\theta$.
On the other hand, the coefficient of the competing cubic term is bounded: $0 \le 4\epsilon_\theta\sin^2\theta \le 4\epsilon_\theta$.
The function $\sin^2\theta$ will oscillate very rapidly compared to $\mu_{\inf}$, and can effectively be replaced by its average value $4\epsilon_\theta\vev{\sin^2\theta} = 2\epsilon_\theta$. The equation of motion is approximately
\begin{equation}
\ddot{\varphi} + \Bigl(3H+\Gamma_\varphi\Bigr)\dot{\varphi} 
-\biggl\{
\dfrac{m_\varphi^2}{2}
+\mu_{\inf}^2\mathcal{A}^2\cos^2(\mu_{\inf}t)
\biggr\}
\,\varphi 
+ \Bigl(\lambda_\phi + 2\epsilon_\theta\Bigr)\varphi^3 \;\approx\; 0\;.
\end{equation}
Therefore, we expect the large oscillating centrifugal barrier term to confine $\varphi$ into either $\varphi > 0$ or $\varphi <0$, soon after the $K$ and $\LCP$ terms become negligible, allowing for a well-defined direction in the evolution of $\theta$.

When $\theta$ is in a phase-locked state, we have $\dot{\theta}\approx n\mu_{\inf}/2$,
and $\sin\theta \approx \sin(n\mu_{\inf}t/2)$.
The equation is,
\begin{equation}
\ddot{\varphi} + \Bigl(3H+\Gamma_\varphi\Bigr)\dot{\varphi} 
-\biggl(
\dfrac{m_\varphi^2}{2}
+\dfrac{n^2\mu_{\inf}^2}{4}
\biggr)
\varphi 
+\biggl\{
\lambda_\phi + 4\epsilon_\theta\sin^2\left(\dfrac{n\mu_{\inf}t}{2}\right)
\biggr\}\varphi^3 \;\approx\; 0\;.
\end{equation}
This time, the cubic term is oscillatory.
The average equilibrium position of $\varphi$ is,
\begin{equation}
\vev{\varphi} 
\;\approx\; \dfrac{n\mu_{\inf}}{2\sqrt{2\epsilon_\theta}}
\;=\; \dfrac{n\mu_{\inf}}{2m_\theta}\,v_\varphi
\;\approx\; \dfrac{\sqrt{n}}{2}\,v_\varphi\;,
\label{vev}
\end{equation}
where we have used Eq.~\eqref{approxn}.
Note that this is the same order of magnitude as $v_\varphi$ 
as assumed, cf. Eqs.~\eqref{ConditionA} and \eqref{ConditionAprime}.
Let $\varphi = \vev{\varphi} + \rho$. 
Assuming $\rho \ll \vev{\varphi} \approx v_\varphi$, the equation for $\rho$ is
\begin{equation}
\ddot{\rho} + \gamma\dot{\rho} 
\;\approx\; \dfrac{\sqrt{n}}{8}\omega^2 v_\varphi\cos\omega t\;,
\end{equation}
where,
\begin{equation}
\gamma \;=\; 3H + \Gamma_\varphi\;,\qquad
\omega \;=\; n\mu_{\inf}\;.
\end{equation}
The amplitude of the forced oscillation of $\rho$ is
\begin{equation}
\alpha 
\;=\; \dfrac{(\sqrt{n}\omega^2 v_\varphi/8)}{\sqrt{\omega^4 + \omega^2\gamma^2}}
\;=\; \dfrac{(\sqrt{n}v_\varphi/8)}{\sqrt{1+(\gamma^2/\omega^2)}}
\quad\xrightarrow{\gamma\ll\omega}\quad \dfrac{\sqrt{n}v_\varphi}{8}
\;,\qquad
\dfrac{\alpha}{\vev{\varphi}}\;\approx\; \dfrac{1}{4}\;.
\end{equation}
Thus, though $\varphi$ can be trapped around $\vev{\varphi} \sim v_\varphi$, its  amplitude of oscillation will 
be of order $\vev{\varphi}/4$. 
%
Therefore, for $\gamma \ll \mu_{\inf}$ the time-evolution of $\rho = \varphi-\vev{\varphi}$ is given by,
\begin{equation}
\rho \;\approx\; \dfrac{1}{4}\vev{\varphi} \cos\omega t\;.
\end{equation}

\bigskip
\subsubsection*{Phase-Locked States Revisited}
\label{PhaseLockedStatesReanalysis}

Let us now return to the reanalysis of the dynamics of $\theta$.
The equation of motion is
\begin{equation}
\ddot{\theta} 
+\left\{
   \Bigl(3H+\Gamma_\theta\Bigr)
   +\left(
       \dfrac{2\dot{\varphi}}{\varphi}
    \right)
 \right\} \dot{\theta} 
\,+\,\epsilon_{\theta}\,\varphi^2\sin 2\theta
\,+\,\dfrac{1}{\LCP}
\left\{
 \mu_{\inf}^2\,\chi
\,-\left(\dfrac{2\dot{\varphi}}{\varphi}\right)\dot{\chi}
\right\}
\;=\; 0
\;.
\label{PL_state}
\end{equation}
The time-dependences of the various terms can be assumed to be given approximately by
\begin{eqnarray}
\chi & = & \chi_i\cos(\mu_{\inf}(t-t_d^0)) ~,\vphantom{\Big|}
\cr
\dot{\chi} & = & -\mu_{\inf}\chi_i\sin(\mu_{\inf}(t-t_d^0)) \vphantom{\Big|}~,
\end{eqnarray}
and
\begin{equation}
\left.
\begin{array}{ll}
\varphi \;=\; \vev{\varphi}\left(1 + \dfrac{1}{4}\cos\left[n \mu_{\inf} (t-t_d^0)\right] \right) \vphantom{\bigg|}
\\
\dot{\varphi} \;=\; -\dfrac{n}{4}\vev{\varphi} \mu_{\inf}\sin(n\mu_{\inf} (t-t_d^0)) \vphantom{\bigg|}
\end{array}
\right\}
\quad\to\quad
\dfrac{2\dot{\varphi}}{\varphi}
\;\approx\; 
-\frac{n}{2}\mu_{\inf}\sin(n\mu_{\inf} (t-t_d^0))
\vphantom{\Big|}~,
\label{PL_eqs}
\end{equation}
where $t_d^0$ is the approximate time that driven motion of $\theta$ becomes possible.
Here, we assume that there is no phase shift between the oscillation of the inflaton $\chi$ and
the induced oscillation of $\varphi$ around $\vev{\varphi}$.

\begin{figure}
	\centering
	\includegraphics[width=0.6\textwidth]{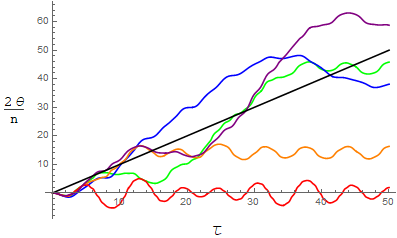}
	\caption{The driven motion observed for different $n$, where $n=2$ (Blue), 3 (Purple), 4 (Green), 4.5 (Orange), 8 (Red) while the black line corresponds to $\dot\theta = (n/2)\mu_{\rm inf}$, and $\tau=\mu_{\rm inf} (t-t_d)$; where we have considered Eq.~\eqref{PL_state} and \eqref{PL_eqs}.}
	\label{driven_motion}
\end{figure}

With these inputs, we solve Eq.~\eqref{PL_state} numerically for the motion of $\theta$, 
and arrive at the solutions depicted in Figure~\ref{driven_motion}. 
Note that in deriving the time dependence of $\varphi$ in the previous section, we assumed that $\theta$
was in a phase-locked state with $\dot\theta = (n/2)\mu_{\inf}$. 
Thus, any driven motion produced in Eq.~\eqref{PL_state} must approximate this relation to be consistent with the analysis assumptions. 
We find that within the range $ 2\le n\le 4 $ the driven motion generated approximates the relation $\dot\theta = (n/2)\mu_{\inf}$ as required. 
For larger values of $n$, consistency with the solution to the equation of motion for $\varphi$ progressively breaks down, 
with $\theta$ showing purely sinusoidal oscillation for $n=8$.

This result can be understood from Eq.~\eqref{vev} and the SSC given in Eq. \eqref{Ch3FinalTimeCondition2}. 
When $n$ is greater than 4, the average value of $\varphi$ given in Eq.~\eqref{vev} becomes greater than 
the value prior to driven motion $ v_\varphi $. The original form of the SSC then quickly becomes violated as the potential of $ \theta $ increases, ending driven motion as visible in the $ n= 4.5 $ and 8 scenarios plotted in Figure \ref{driven_motion}.
Instead for $n\leq 4$, the average value of $\varphi$ decreases and subsequently lowers the required driving force to maintain the SSC. This leads to the continued production of driven motion in $ \theta $ as seen in the $ n=2,~3, $ and 4 cases depicted in Figure \ref{driven_motion}.

\subsection{Higgs Vacuum Stability}

To ensure the positivity of $\lambda_h(\mu)$ up to the Planck scale, 
we must demand the mass ordering $m_\varphi < \mu_{\mathrm{ins}} < \mu_{\inf} < m_\theta$,
as well as the condition given in Eq.~\eqref{kappa-bound2}, namely $\kappa^2 > 0.04\,\lambda_\phi$.
Here, we treat $\kappa$ and $\lambda_\phi$ as scale independent objects since their RGE running
can be expected to be negligible due to the small sizes they will be assigned.

Combining this constraint with that in Eq.~\eqref{ConditionK} we find the range of $\kappa$ consistent with our analysis to be
\begin{equation}
0.04\,\lambda_\phi \,<\, \kappa^2 \,\ll\, 24\left(\dfrac{\mu_{\inf}}{v_\varphi}\right)^4 \;,
\label{kappa-range}
\end{equation}
which implies
\begin{equation}
0.0004
\biggl(
  \dfrac{2\lambda_\phi v_\varphi^2}{\mu_{\inf}^2}
\biggr)^2
=\; 0.0004
\left(\frac{m_\varphi}{\mu_{\inf}}\right)^4
\ll\;\lambda_\phi\;.
\label{Condition3}
\end{equation}
Since we need to require $m_\varphi = \sqrt{2\lambda_\phi} v_\varphi < \mu_\mathrm{ins} = 10^{10}\,\mathrm{GeV}$ 
for Higgs vacuum stability, this condition is not difficult to satisfy.

\subsection{Generated Baryon Asymmetry}

Now that we have found the parameter requirements to control the dynamics of $ \varphi $ and $ \chi $, and thus produce  driven motion, we can follow the Pendulum Leptogenesis mechanism analysis as described in Sec.~\ref{Dym_mod} to calculate the generated baryon asymmetry. In this scenario, the UV cut-off scale will be fixed to,
\begin{equation}
\Lambda_{\cancel{\mathcal{CP}}} \;=\; \frac{M_p}{\zeta}\;.
\end{equation}
where this is motivated by the scale at which the Einstein and Jordan Frame descriptions of $ \varphi $ coincide, allowing the consistent definition of the $ \mathcal{CP} $ violating interaction between the Higgs and $ \varphi $, as given in Eq. \eqref{CP_term}, during the reheating epoch.

In this setting, the Sweet Spot Condition has the following form,
\begin{equation}
n \;\simeq\; 5\frac{\zeta H_d}{\mu} 
\quad\mbox{or}\quad 
H_d \;\simeq\;\frac{n\mu}{5\zeta}
\;.
\end{equation}
and from the initial Hubble rate of reheating, Eq. \eqref{ReheatingHubbleRange}, we  obtain the allowed range,
\begin{equation}
\zeta \;>\; \frac{n}{2} \;\ge\; 1\;,
\end{equation}
while from the analysis of $ \varphi $, we require $ 4 \;\ge\; n\;\ge\; 2\; $ for a consistent description of the driven motion, so we can simply require $ \zeta>2 $~. The $ \zeta $ coupling has an upper bound from  the requirement in Eq. \eqref{ConditionAprime}, and thus $ \zeta $ must exist within the range,
\begin{equation}
\frac{M_p }{v_{\varphi}} \;>\; \zeta \;>\; 2\;.
\label{zeta_range}
\end{equation}

Successful generation of driven motion leads to a baryon asymmetry parameter analogous to that found earlier in Eq. \eqref{etaB},
\begin{equation}
\eta_B 
\;=\; \frac{28}{79}\,\eta_L 
\;\simeq\;
0.18\;\frac{\epsilon_{\theta} v_{\varphi}^4 T_{\mathrm{reh}}}{\mu_{\inf} H_d^2 M_p^2} 
\;.
\end{equation}
Therefore, in this scenario the observed baryon asymmetry is given by,
\begin{eqnarray}
\frac{\eta_B}{\eta_B^\mathrm{obs}} 
\;\simeq\; 
\frac{2\zeta^2}{3\epsilon_{\theta} }\left(\frac{T_{\mathrm{reh}}}{2\times 10^{13} \textrm{~GeV}}\right)
\;,
\end{eqnarray}
where we have utilised the SSC to rearrange this equation. Considering the range given in Eq. \eqref{zeta_range} and the approximate reheating temperature for Higgs Inflation we have the following bound on the asymmetry generation,
\begin{eqnarray}
\frac{2 M_p}{3\epsilon_{\theta} v_\varphi }>\frac{\eta_B}{\eta_B^\mathrm{obs}} 
>
\frac{8}{3\epsilon_{\theta} }
\;,
\end{eqnarray}
 seeing as we require $ \epsilon_{\theta}<1 $ an approximate lower limit on the asymmetry can be given by, 
\begin{eqnarray}
\frac{\eta_B^{\rm min}}{\eta_B^\mathrm{obs}} \gtrsim \mathcal{O}(10)
\;,
\end{eqnarray}
which is promising given that we have assumed ideal conditions.

\subsection{Summary of Scales Involved}
The parameter constraints in the $ \zeta,\kappa \neq 0 $ scenario are different to those summarised in Section \ref{SSI}, due to the presence of these new couplings and the requirement of Higgs vacuum stability. Here, the ordering of the mass parameters has changed to,
\begin{eqnarray}
m_\varphi \;\leq\; \mu_{\rm ins}\;<\; \mu_{\inf}  &\;\le\;& m_\theta   \cr
& \downarrow & \cr
\sqrt{\lambda_\phi}  & < & \sqrt{\epsilon_\theta}\cr
& \downarrow & \cr
\lambda_\phi  & \ll & \epsilon_\theta
\label{MassOrdering2}
\end{eqnarray}
to enable the scalar $ \varphi $ to bring stability to the Higgs vacuum. The subsequent requirements on the parameter  that achieve this are, 
\begin{equation}
\kappa^2 > 0.04\,\lambda_\phi
~~~~\textrm{and}~~~~
 0.0004
\left(\frac{m_\varphi}{\mu_{\inf}}\right)^4
\ll\;\lambda_\phi\;,
\label{Condition}
\end{equation}
ensuring $ \lambda_h $ is positive up to the Planck scale. These can all be satisfied consistently with the Pendulum Leptogenesis scenario and successful Higgs Inflation.

The allowed range of $ n $ is reduced due to the inclusion of the dynamics of $ \varphi $. For a consistent description of the driven motion we require, 
\begin{equation}
 4 \;\ge\; n\;\ge\; 2\;, 
\end{equation}
which places a lower bound on the non-minimal coupling $ \zeta $ through the cut-off scale $\LCP$.
In this scenario, $\LCP$ is defined by the energy scale at which the Einstein and Jordan frame fields of $ \varphi $ converge,
\begin{equation}
\Lambda_{\cancel{\mathcal{CP}}} \;=\; \frac{M_p}{\zeta}\;, 
\end{equation}
which means that $ \zeta $ must lie in the range, 
\begin{equation}
\frac{M_p }{v_{\varphi}} \;>\; \zeta \;>\; 2\;.
\label{zeta_range1}
\end{equation}

Selecting parameters satisfying each of these conditions lead to successful Higgs inflation, Pendulum Leptogenesis during reheating, and stability of the Higgs vacuum. A possible set of parameters which satisfies all these constraints is,
\begin{eqnarray}
\LCP & = & 3\times 10^{16}\,\mathrm{GeV}\;,\cr
v_\varphi & = & 10^{15}\,\mathrm{GeV}\;,\cr
\epsilon_\theta & \simeq & 2\times 10^{-3}\;,\cr
\lambda_\phi & = & 5\times 10^{-11}\;,\cr
\lambda_h\left(V_{\inf}^{1/4}\,\right) & \simeq & 0.07\;,\cr
\kappa & = & 4\times 10^{-6}\;,\cr
\xi & \simeq & 1.2\times 10^4\;,\cr
\zeta & = & 80\;,\cr
n & = & 4\;, 
\label{parameters}
\end{eqnarray}
which corresponds to
\begin{eqnarray}
\mu_{\inf} & = & 3\times 10^{13}\,\mathrm{GeV}\;,\cr
m_\theta \,& \simeq & 6\times 10^{13}\,\mathrm{GeV}\;,\cr
m_\varphi & = &  10^{10}\,\mathrm{GeV}\;,\cr
T_{\mathrm{reh}} & \simeq & 2.7\times 10^{13}\,\mathrm{GeV}\;,
\end{eqnarray}
and
\begin{equation}
\dfrac{\eta_B}{\eta_B^{\mathrm{obs}}}
\;\approx\; 3 \times 10^6\;.
\end{equation}
The value of $\lambda_h$ given in Eq. (\ref{parameters}) is evaluated at the inflationary scale, $ V_{\inf}=\dfrac{3}{4}\mu_{\inf}^2 M_p^2 $, taking into account the scalar threshold effect given in Eq. (\ref{thresh}). The Higgs non-minimal coupling $ \xi $ is then fixed by combining this result with observation in Eq. (\ref{lambda_over_xi2}).

\section{Conclusion}
\label{Conclusion}

We have presented a model that entails a minimal addition to the Standard Model to 
simultaneously ensure Higgs vacuum stability up to the Planck scale, successful inflation, Leptogenesis via the pendulum mechanism, and generate the active neutrino masses. 
Considering the two components of the complex scalar lepton, the real scalar and complex phase, can help illuminate how this is achieved. The real scalar component couples to the Higgs boson via the portal interaction and to gravity via a non-minimal coupling, meaning that it can take part in the inflationary epoch and ensure vacuum stability, while giving a mass to the right handed neutrinos via its vacuum expectation value. 
On the other hand, the complex scalar phase is integral to the setup of the pendulum dynamics and hence the Leptogenesis mechanism during reheating. 
This model was found to simultaneously achieve the goals of Higgs vacuum stability, successful inflation, Leptogenesis via the pendulum mechanism, and light neutrino masses. The parameter conditions required to achieve this were explored and summarised.

\section*{Acknowledgements}

We would like to thank James Gray for helpful discussions.
NDB is supported by the World Premier International Research Center Initiative (WPI), MEXT, Japan. 
TT is supported in part by the US Department of Energy (DE-SC0020262) 
and by the US National Science Foundation (NSF Grant 1413031).
KY is supported by the Chinese Academy of Sciences (CAS) 
President's International Fellowship Initiative under Grant No. 2020PM0018.
KY's work was also supported in part by the National Center for Theoretical Sciences, Taiwan.

\newpage
\appendix
\section{Minima of the Effective Quartic Coupling $\lambda_\mathrm{eff}(s_\delta^2)$}
\label{AppLambdaEffMin}

We wish to find the minimum of the function
\begin{eqnarray}
\lambda_\mathrm{eff}(s_\delta^2) 
& = & A c_\delta^4 + B c_\delta^2 s_\delta^2 + C s_\delta^4
\vphantom{\Big|}\cr
& = & A(1-s_\delta^2)^2 + B(1-s_\delta^2)s_\delta^2 + C s_\delta^4 
\vphantom{\Big|}\cr
& = & (A-B+C)s_\delta^4 - (2A-B)s_\delta^2 + A
\vphantom{\Big|}\cr
& = & (A-B+C)
\left[s_\delta^2 - \dfrac{A-(B/2)}{A-B+C}\right]^2
+ \dfrac{AC-(B^2/4)}{A-B+C}
\;,
\end{eqnarray}
where $s_\delta=\sin\delta$, $c_\delta=\cos\delta$. 
Note that the range of $s_\delta^2$ is $0\le s_\delta^2\le 1$.
We will assume $A>0$, $C>0$, but allow $B$ to be of either sign.
Note that the stability of the potential demands
\begin{equation}
4AC \,>\, B^2
\quad\to\quad -2\sqrt{AC}\,<\,B\,<\,2\sqrt{AC}
\;,
\label{BrestrictionApp}
\end{equation}
cf. Eq.~\eqref{Vstability}.

\begin{itemize}
\item If $(A-B+C)<0$ then $\lambda_\mathrm{eff}(s_\delta^2)$ is a concave quadratic function of $s_\delta^2$, 
and its minimum in the range $0\le s_\delta^2\le 1$ is at either $s_\delta^2=0$ or $s_\delta^2=1$:
\begin{equation}
\lambda_\mathrm{eff}(0)\,=\,A
\,,\qquad 
\lambda_\mathrm{eff}(1)\,=\,C
\,.
\end{equation}
So if $A+C<B$ and $A<C$, then the minimum is $A$ at $s_\delta^2=0$
and if $A+C<B$ and $A>C$, then the minimum $C$ is at $s_\delta^2=1$.

\item If $(A-B+C)>0$ then $\lambda_\mathrm{eff}(s_\delta^2)$ is a convex quadratic function of $s_\delta^2$. 
The lowest point of the parabola is at
\begin{equation}
s_\delta^2 \;=\; \dfrac{A-(B/2)}{A-B+C}\;,
\end{equation}
but this does not always fall within the interval $s_\delta^2\in[0,1]$.
There are three possible cases:
\begin{equation}
\begin{array}{ll}
\dfrac{A-(B/2)}{A-B+C}\,<\,0
&\quad\to\quad
A<C,\;\;2A<B<A+C
\quad\to\quad
\mbox{minimum $A$ at $s_\delta^2 = 0$} 
\\
1\,<\,\dfrac{A-(B/2)}{A-B+C}
&\quad\to\quad
C<A,\;\;2C<B<A+C
\quad\to\quad
\mbox{minimum $C$ at $s_\delta^2 = 1$} 
\\
0\,<\,\dfrac{A-(B/2)}{A-B+C}\,<\,1
&\quad\to\quad
B<\min(2A,2C,A+C)
\\
&\quad\to\quad
\mbox{minimun $\dfrac{AC-(B^2/4)}{A-B+C}$ at
$s_\delta^2 = \dfrac{A-(B/2)}{A-B+C}$}
\end{array}
\end{equation}
Note that Eq.~\eqref{BrestrictionApp} ensures the positivity of 
$\lambda_\mathrm{eff}(s_\delta^2)$ in the third case.

\end{itemize}

\noindent
In the main text, we assume $A<C$, in which case
if $2A<B<2\sqrt{AC}$ the minimum is $A = \lambda_h/\xi^2$ at $s_\delta^2=0$,
while if $-2\sqrt{AC}<B<2A$ the minimum is
\begin{equation}
\dfrac{AC-(B^2/4)}{A-B+C}
\;=\;
\dfrac{\dfrac{\lambda_h\lambda_\phi}{\xi^2\zeta^2}-\dfrac{\kappa^2}{4\xi^2\zeta^2}}
{\dfrac{\lambda_h}{\xi^2}-\dfrac{\kappa}{\xi\zeta}+\dfrac{\lambda_\phi}{\zeta^2}}
\;=\;
\dfrac{\lambda_h\lambda_\phi - (\kappa^2/4)}
      {\zeta^2\lambda_h - \xi\zeta\kappa + \xi^2\lambda_\phi}
\;\le\;\dfrac{\lambda_h}{\xi^2}\;<\;\dfrac{\lambda_\phi}{\zeta^2}\;.
\end{equation}
at
\begin{equation}
s_\delta^2 \;=\; \dfrac{A-(B/2)}{A-B+C}
\;=\; 
\dfrac{\dfrac{\lambda_h}{\xi^2}-\dfrac{\kappa}{2\xi\zeta}}
      {\dfrac{\lambda_h}{\xi^2}-\dfrac{\kappa}{\xi\zeta}+\dfrac{\lambda_\phi}{\zeta^2}}
\;=\;
\dfrac{\zeta^2\lambda_h - \xi\zeta\kappa/2}{\zeta^2\lambda_h - \xi\zeta\kappa + \xi^2\lambda_\phi}
\;.
\end{equation}
Therefore, even when $\zeta\neq 0$, the inflaton will consist of only the Higgs
($s_\delta^2=0$)
if the conditions
\begin{equation}
A<C,\;\;2A<B<2\sqrt{AC},
\quad\to\quad
\dfrac{\lambda_h}{\xi^2} < \dfrac{\lambda_\phi}{\zeta^2}\;,\quad
\dfrac{2\zeta\lambda_h}{\xi} < \kappa < 2\sqrt{\lambda_h\lambda_\phi}\;,
\end{equation}
are met. If
\begin{equation}
A<C,\;\;-2\sqrt{AC}<B<2A
\quad\to\quad
\dfrac{\lambda_h}{\xi^2} < \dfrac{\lambda_\phi}{\zeta^2}\;,\quad
-2\sqrt{\lambda_h\lambda_\phi} < \kappa < \dfrac{2\zeta\lambda_h}{\xi}\,,
\end{equation}
the inflaton will be a mixture of $h$ and $\varphi$.
The maximum value of $s_\delta^2$ where the potential minimum is is given by
\begin{equation}
s_\delta^2 \;=\; \dfrac{\sqrt{A}}{\sqrt{A}+\sqrt{C}}
\end{equation}
when $B=-2\sqrt{AC}$, that is $\kappa = -2\sqrt{\lambda_h\lambda_\phi}$, at which point $\lambda_\mathrm{eff}(s_\delta^2)$ 
will be zero, and go negative if $B$ is decreased further into the negative.
Therefore, if $A\ll C$, the inflaton $\chi$ will always be dominated by $h$.

\newpage
\section{2-loop RGEs for Standard Model plus Lepton Scalar Portal}
\label{2loopRGE}

In our analysis we utilize the following 1-loop and 2-loop RGE coefficients obtained from Ref.~\cite{Salvio:2018rv}.
Here, $g_1$, $g_2$, $g_3$ are respectively the $U(1)_Y\times SU(2)_L\times SU(3)$ gauge couplings,
$y_t$ is the top Yukawa, and 
$\beta_{g}^{(n)}/(4\pi)^{2n}$ are the $n$-loop contributions to the $\beta$-function of $g$ above the scale $m_\varphi$.
Below the scale $m_\varphi$, the $\lambda_\phi$ and $\kappa$ terms are absent.

\begin{itemize}
\item 1-loop:
\begin{eqnarray}
\beta_{g_1^2}^{(1)} & = &   \frac{41g_1^4}{10}\;, \qquad   
\beta_{g_2^2}^{(1)} \,=\, - \frac{19g_2^4}{6}\;, \qquad
\beta_{g_3^2}^{(1)} \,=\, -\frac{19 g_3^4}{3}\;,
\cr
\beta_{y_t^2}^{(1)} & = & y_t^2\left(\frac92 y_t^2-8g_3^2-\frac{9g_2^2}{4}-\frac{17g_1^2}{20}\right),
\cr
\beta_{\lambda_h}^{(1)} & = & \lambda_h\left(12\lambda_h+6y_t^2-\frac{9g_1^2}{10}-\frac{9g_2^2}{2}\right) - 3y_t^4 +\frac{9 g_2^4}{16}+\frac{27 g_1^4}{400}+\frac{9 g_2^2 g_1^2}{40}+\frac{\kappa^2}{2} ,
\cr
\beta_{\kappa}^{(1)} & = & \kappa\left(3y_t^2-\frac{9g_1^2}{20}-\frac{9g_2^2}{4}+6\lambda_h \right) +4\lambda_\phi  \kappa+2 \kappa^2, 
\cr
\beta_{\lambda_\phi}^{(1)} & = & \kappa^2+10\lambda_\phi^2\;.
\end{eqnarray} 
%

\item 2-loop:
\begin{eqnarray}
\beta_{g_1^2}^{(2)} & = & g_1^4 \left(\frac{199 g_1^2}{50}+\frac{27 g_2^2}{10}+\frac{44 g_3^2}{5}-\frac{17 y_t^2}{10}\right), 
\cr
\beta_{g_2^2}^{(2)} & = & g_2^4 \left(\frac{9 g_1^2 }{10}+\frac{35 g_2^2}{6}+12  g_3^2-\frac{3  y_t^2}{2}\right),
\cr
\beta_{g_3^2}^{(2)} & = & g_3^4 \left(\frac{11 g_1^2}{10}+\frac{9g_2^2}{2}-\frac{40 g_3^2}{3}-2 y_t^2\right), 
\cr
\beta_{y_t^2}^{(2)} & = & y_t^2 \bigg[ 6\lambda_h^2 -\frac{23 g_2^4}{4}+
y_t^2 \left( -12 y_t^2 -12 \lambda_h 
+36 g_3^2+\frac{225 g_2^2}{16}+\frac{393 g_1^2}{80} 
\right) 
\cr & & \quad
+\,\frac{1187 g_1^4}{600}+9 g_3^2 g_2^2
\,+\,\frac{19}{15} g_3^2 g_1^2
\,-\,\frac{9}{20} g_2^2 g_1^2 
\,-\,\frac{932 g_3^4}{9}
\,+\,\frac{\kappa^2}{2}
\bigg],
\cr
\beta_{\lambda_h}^{(2)} & = & \lambda_h^2 \left[54 \left(g_2^2+\frac{g_1^2}{5}\right) \right. -156\lambda_h -72 y_t^2 \bigg] 
+\lambda_h y_t^2 \left( 40 g_3^2
+\frac{45 g_2^2}{4}+\frac{17 g_1^2}{4}-\frac32 y_t^2\right)
\cr & &
+\lambda_h \bigg(\frac{1887 g_1^4}{400} -\frac{73 g_2^4}{16}
+\frac{117 g_2^2 g_1^2}{40}
-5 \kappa^2\bigg) + y_t^4 \left( 15 y_t^2-16g_3^2-\frac{4 g_1^2}{5}\right)
-2\kappa^3
\cr & &
+y_t^2 \left(\frac{63 g_2^2g_1^2}{20} -\frac{9 g_2^4}{8}-\frac{171 g_1^4}{200}\right)  +\frac{305 g_2^6}{32} -\frac{3411 g_1^6}{4000} -\frac{289 g_2^4 g_1^2}{160} -\frac{1677 g_2^2 g_1^4}{800}  
\;.
\cr & &
\end{eqnarray}

\end{itemize}

\newpage
\bibliographystyle{JHEP}
\bibliography{Higgs_pendulum}

\end{document}